\documentclass{aa}
\usepackage{graphics,psfig}

\newcommand{\ratio} {N({\rm H}_2) / I_{\rm CO(1 \rightarrow 0)}}
\def\la{\lower.5ex\hbox{$\; \buildrel < \over \sim \;$}}
\def\ga{\lower.5ex\hbox{$\; \buildrel > \over \sim \;$}}

\begin{document}      

   \title{New CO observations and simulations of the NGC~4438/NGC~4435 system \thanks{Movies
     of all simulations are available in the electronic version of this article.}}

   \subtitle{Interaction diagnostics of the Virgo cluster galaxy NGC~4438}

   \author{B.~Vollmer\inst{1,2}, J. Braine\inst{3}, F.~Combes\inst{4}, \& Y.~Sofue\inst{5}}

   \offprints{B.~Vollmer, e-mail: bvollmer@astro.u-strasbg.fr}

   \institute{CDS, Observatoire astronomique de Strasbourg, UMR 7550, 11, rue de l'universit\'e, 
     67000 Strasbourg, France \and
     Max-Planck-Institut f\"ur Radioastronomie, Auf dem H\"ugel 69, 53121 Bonn Germany \and
     Observatoire de Bordeaux (OASU), UMR 5804, CNRS/INSU, B.P. 89, F-33270 Floirac, France \and
     Observatoire de Paris, LERMA, 61 Av. de l'Observatoire, 75014 Paris, France \and
     Institute of Astronomy, University of Tokyo, Mitaka, Tokyo 181-0015, Japan
   } 
          
   \date{Received / Accepted}

   \authorrunning{Vollmer et al.}
   \titlerunning{Interaction diagnostics of NGC~4438}

\abstract{
NGC~4438 is a highly perturbed spiral with a stellar tidal tail and extraplanar 
molecular gas, now very HI deficient, crossing the center of the Virgo cluster 
at high speed.  Various authors have attributed the perturbed appearance to the
ram pressure of the intracluster medium, the tidal interaction with NGC~4435,
and an ISM-ISM collision between the ISM of NGC~4438 and NGC~4435.  
We present new CO observations covering virtually all of NGC~4438 and the center
of NGC~4435 and detailed simulations including all of the above effects.
For the first time CO is detected in NGC~4435.
In NGC~4438 we find double line profiles at distances up to $40''$ to the west and south-west
and redshifted lines with respect to galactic rotation in the south of the center.
The lack of gas to the North and East coupled with the large gaseous extent to the West and
the redshifted and double line profiles can only be reproduced with a ram pressure wind.
NGC~4438 is most probably on its first passage 
through the cluster center and has been stripped of its HI only over the past 
100~Myr. While an ISM-ISM collision between NGC~4435 and NGC~4438 may occur, 
the effect is not significant compared to ram pressure and tidal forces,
not surprising for the passage of an S0 galaxy $5 - 10$~kpc from the center 
of NGC 4438.  We also detect narrow CO lines, in the absence of detected HI, in the 
northern tidal arm some 15~kpc from the center of NGC~4438.  This can be understood 
from the simulations assuming a few percent of the gas is too dense to experience the ram pressure 
wind. NGC~4438 has changed greatly over the past 100~Myr due to its plunge through 
the center of the Virgo cluster  and the interaction with the S0 NGC~4435.
The ram pressure wind has a strong effect which is increased by the interaction with NGC~4435, 
bringing gas further from NGC~4438 where the ram pressure strips it away.
}

\maketitle

\section{Introduction}

\begin{figure}
        \resizebox{\hsize}{!}{\includegraphics{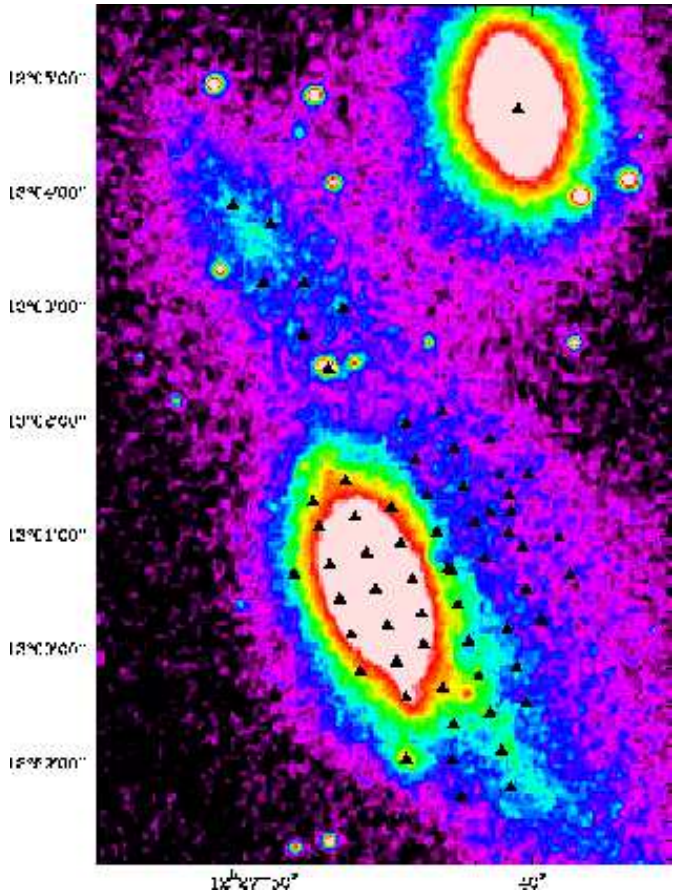}}
        \caption{Optical B band image of the NGC~4438/NGC~4435 system.
	The observed positions are marked with triangles.
        } \label{fig:optical}
\end{figure}

The perturbed spiral galaxy NGC~4438 and its companion galaxy NGC~4435 represent the most complicated 
system in the Virgo cluster (see Fig.~\ref{fig:optical} and Table~\ref{tab:parameters}). 
The highly inclined disk of NGC~4438 is heavily perturbed showing a
prominent tidal perturbation to the north and with stellar debris displaced to the west of the galaxy's 
main disk (Arp 1966). Combes et al. (1988) simulated the NGC~4438/NGC~4435 system using a test-particle code.
They showed that the northern tidal tail can be reproduced by a retrograde encounter with an impact
parameter of $\sim 6$~kpc that occurred $\sim 100$~Myr ago. The relative position and radial velocity 
between NGC~4438 and NGC~4435 of 730~km\,s$^{-1}$ could be reproduced by their model.

Observations of the interstellar medium (ISM) of NGC~4438 revealed an important extraplanar
component that is displaced to the west of the galactic disk (for an overview of the gas distribution
at multiple wavelengths see Fig.11 of Kenney et al. 1995). The disk ISM is mainly in the form
of molecular gas (Combes et al. 1988, Kenney et al. 1995) characterized by the absence of H{\sc i}
emission (Cayatte et al. 1990, Hibbard et al. 2001). 
However, CO and H{\sc i} are detected to the west of the center
of NGC~4438. The most spectacular gas distribution is that of the dense ionized gas observed in H$\alpha$
(Kenney et al. 1995). These observations revealed several filaments which originate from the disk
plane and extend out from the disk for 5-10~kpc towards the west and southwest. The radial velocities
of these filaments are within $\sim 200$~km\,s$^{-1}$ of the galaxy's systemic velocity.
Not only the cold and warm phases of the ISM are detected to the west beyond the galactic disk,
but also the hot phase (X-rays) and the magnetic field (radio continuum; Kotanyi et al. 1983). 

While it is certain that the distortion of the stellar content of NGC~4438 is due to
a tidal encounter with NGC~4435, different mechanisms were put forward to explain the displacement of 
all ISM phases: (i) the tidal interaction which extracted the molecular gas from the center
and left it to the west of the galaxy (Combes et al. 1988), (ii) a collision between the
ISM of the two galaxies (Kenney et al. 1995), and (iii) ram pressure stripping due to the rapid
motion of NGC~4438 through the hot intracluster medium (ICM; Kotanyi et al. 1983, Chincarini
\& de Souza 1985). 

Since NGC~4438 is located in projection only 1$^{\rm o} \sim 300$~kpc from the cluster 
center\footnote{We use a distance to the Virgo cluster of 17~Mpc} and has a radial velocity of 
$\sim 1000$~km\,s$^{-1}$ with respect to the cluster mean, the conditions necessary for
strong ram pressure are fulfilled. On the other hand, since Kenney et al. (1995) detected
H$\alpha$ and Machacek et al. (2004) detected soft X-ray emission in the center of NGC~4435, 
an ISM-ISM collision is unavoidable if
the impact parameter of the encounter with NGC~4438 is smaller than $\sim 10$~kpc. 
On the observational side, extraplanar CO and extraplanar, asymmetric, high surface brightness
radio continuum emission are very rare phenomena. 
It is mainly observed in direct galaxy collisions where the two ISM collide
(see e.g. Braine et al. 2003). On the other hand, displacements of X-ray 
and $H\alpha$ emission are observed in cluster spiral galaxies (X-rays: Finoguenov et al. 2004, 
in preparation for the Coma cluster; H$\alpha$: Yoshida et al. 2002; NGC~4388 in the Virgo cluster).
Thus, there is no clear evidence that rules out or confirms one of the suggested interactions.

In this article we present new CO observations of NGC~4438 and NGC~4435 (Sect.~\ref{sec:observations})
together with numerical simulations of the system (Sect.~\ref{sec:simulations}). 
The latter present the advantage of separating different types of interaction to determine their 
influence on the ISM. The CO observations show three characteristics that can only be reproduced 
by a model including ram pressure stripping (Sect.~\ref{sec:doubles}): 
(i) the extraplanar gas distribution, (ii) extraplanar double line profiles, and (iii) redshifted
line profiles with respect to galactic rotation.

\begin{table}
      \caption{Physical parameters of NGC~4438}
         \label{tab:parameters}
      \[
         \begin{array}{lr}
            \hline
            \noalign{\smallskip}
        {\rm Other\ names} &  {\rm Arp~120} \\
                & {\rm VCC~1043} \\
	        & {\rm UGC~7574} \\
        $$\alpha$$\ (2000)$$^{\rm a}$$ &  12$$^{\rm h}27^{\rm m}45.91^{\rm s}$$\\
        $$\delta$$\ (2000)$$^{\rm a}$$ &  +13$$^{\rm o}00'32.3''$$\\
        {\rm Morphological\ type}$$^{\rm a}$$ & {\rm Sb} \\
        {\rm Distance\ to\ the\ cluster\ center}\ ($$^{\rm o}$$) & 1.0\\
        {\rm Optical\ diameter\ D}_{25}$$^{\rm a}$$\ ($$'$$) & 8.5\\
        {\rm B}$$_{T}^{0}$$$$^{\rm a}$$ & 10.49\\ 
        {\rm Systemic\ heliocentric\ velocity}$$^{\rm a}$$\ {\rm (km\,s}$$^{-1}$$)\ & 71$$\pm$$3\\
        {\rm Distance\ D\ (Mpc)} & 17 \\
        {\rm PA} & 29$$^{\rm o}$$\ $$^{\rm b}$$\\
        {\rm Inclination\ angle} & 80$$^{\rm o}$$\ $$^{\rm b}$$,\ 85$$^{\rm o}$$\ $$^{\rm c}$$\\
        {\rm HI\ deficiency}^{\rm d}$$ &  >1.0\\
        \noalign{\smallskip}
        \hline
        \end{array}
      \]
\begin{list}{}{}
\item[$^{\rm{a}}$] RC3, de Vaucouleurs et al. (1991)
\item[$^{\rm{b}}$] Kenney et al. (1995)
\item[$^{\rm{c}}$] Combes et al. (1988)
\item[$^{\rm{d}}$] Cayatte et al. (1994)
\end{list}
\end{table}

\section{Observations \label{sec:observations}}

The observations were carried out with the 30~meter millimeter-wave
telescope on Pico Veleta (Spain) run by the Institut de RadioAstronomie
Millim\'etrique (IRAM) in June 2002.  The CO(1--0) and
CO(2--1) transitions at 115 and 230~GHz respectively were
observed simultaneously and in both polarizations. 
The observed positions are marked as triangles in Fig.~\ref{fig:optical}. 
We generally used the $512 \times 1$~MHz filterbanks at 3mm and 
the two $256 \times 4$~MHz filterbanks at 1mm, yielding an 
instantaneous bandwidth of 1300~km\,s$^{-1}$.
 
System temperatures were typically 250--350~K at 3mm and much higher
for the CO(2--1) transition (T$_A^*$ scale), due to the high water 
vapor content (usually H$_2$O $>$ 6mm).  The forward (main beam)
efficiencies at Pico Veleta were taken to be 0.95 (0.74) at
115~GHz and 0.90 (0.54) at 230~GHz.  
The half-power beamwidths are about 21$''$ and 11$''$.
All observations were done in wobbler-switching mode, usually with a
throw of 150$''$ but sometimes more or less depending on the position
observed, in order to be sure not to have emission in the reference beam.
Pointing was checked on the bright quasar 3C273 roughly every 90 minutes.  

The main observational problem was the anomalous refraction
that affected pointing measurements and resulted in a widening of the beam
until sunset.  Much of the badly affected data were rejected and we
do not present the CO(2--1) data due to the generally high noise 
level and refraction.
 
Data reduction was very simple.  After eliminating the 
bad spectra or bad channels, the spectra for each position were
summed.  Only zero-order baselines (i.e. continuum levels) were
subtracted to obtain the final spectra. 

Two features are particularly interesting: the presence of molecular gas in the
absence of H{\sc i} in the tidal tail to the north (Cayatte et al. 1990,
Hibbard et al. 2001) and the molecular gas to the west initially detected by Combes et al. (1988).

\subsection{NGC~4435 \label{sec:n4435}}

\begin{figure}
	\psfig{file=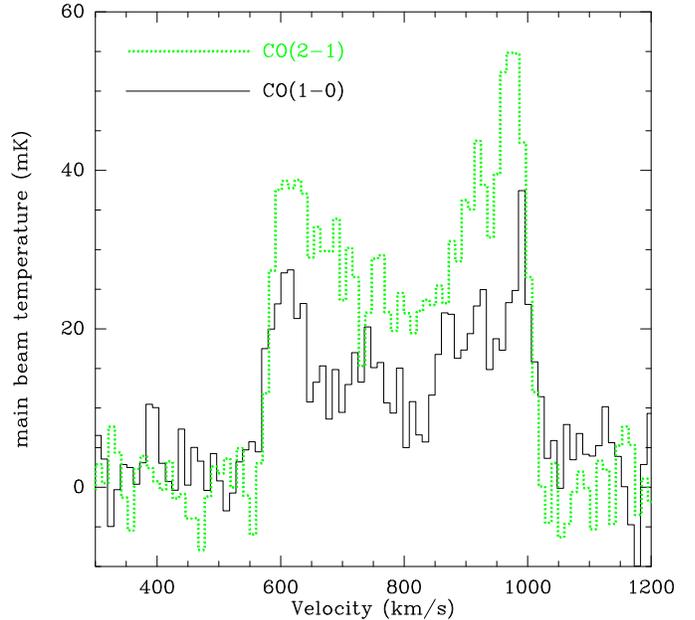,angle=-90,width=\hsize}
        \caption{CO(1--0) and CO(2-1) spectra of NGC~4435.
	The systemic velocity of NGC~4435 is 800~km\,s$^{-1}$.
        } \label{fig:n4435}
\end{figure}

We made one pointing on the center of NGC~4435. For the first time
CO emission is detected in this galaxy.
The CO(1--0) and CO(2--1) spectra are shown in Fig.~\ref{fig:n4435}. 
The total line width is $\sim 400$~km\,s$^{-1}$, which is about $100$~km\,s$^{-1}$
larger than the line width derived from optical spectroscopy (Kenney et al. 1995).
This large linewidth justifies our model rotation velocity for NGC~4435
(see Sect.~\ref{sec:simulations}). In spiral galaxies the ratio between the CO(2--1)
and CO(1--0) lines observed at the same resolution is about 0.9 (Braine et al. 1993). 
Since the brightness temperature of an unresolved source is inversely proportional to
the beamsize, a larger line ratio than 0.9 implies that the source is not
resolved at 115~GHz. The fact that the ratio between the CO(2--1)
and CO(1--0) lines is about two indicates that the spatial extent of the
molecular gas is comparable to the CO(2--1) resolution of $11'' \sim 900$~pc,
somewhat larger than the extent estimated by Kenney et al. (1995). 
Using the same conversion factor of 
$\ratio = 2 \times 10^{20}$~H$_2$~cm$^{-2}$ (K km/s)$^{-1}$, the molecular gas 
mass of NGC 4435 is $9.5 \times 10^{7}$~M$_{\odot}$ 
when the He associated with the molecular gas is included.

\subsection{NGC~4438 \label{sec:spectra}}

\begin{figure*}
	\psfig{file=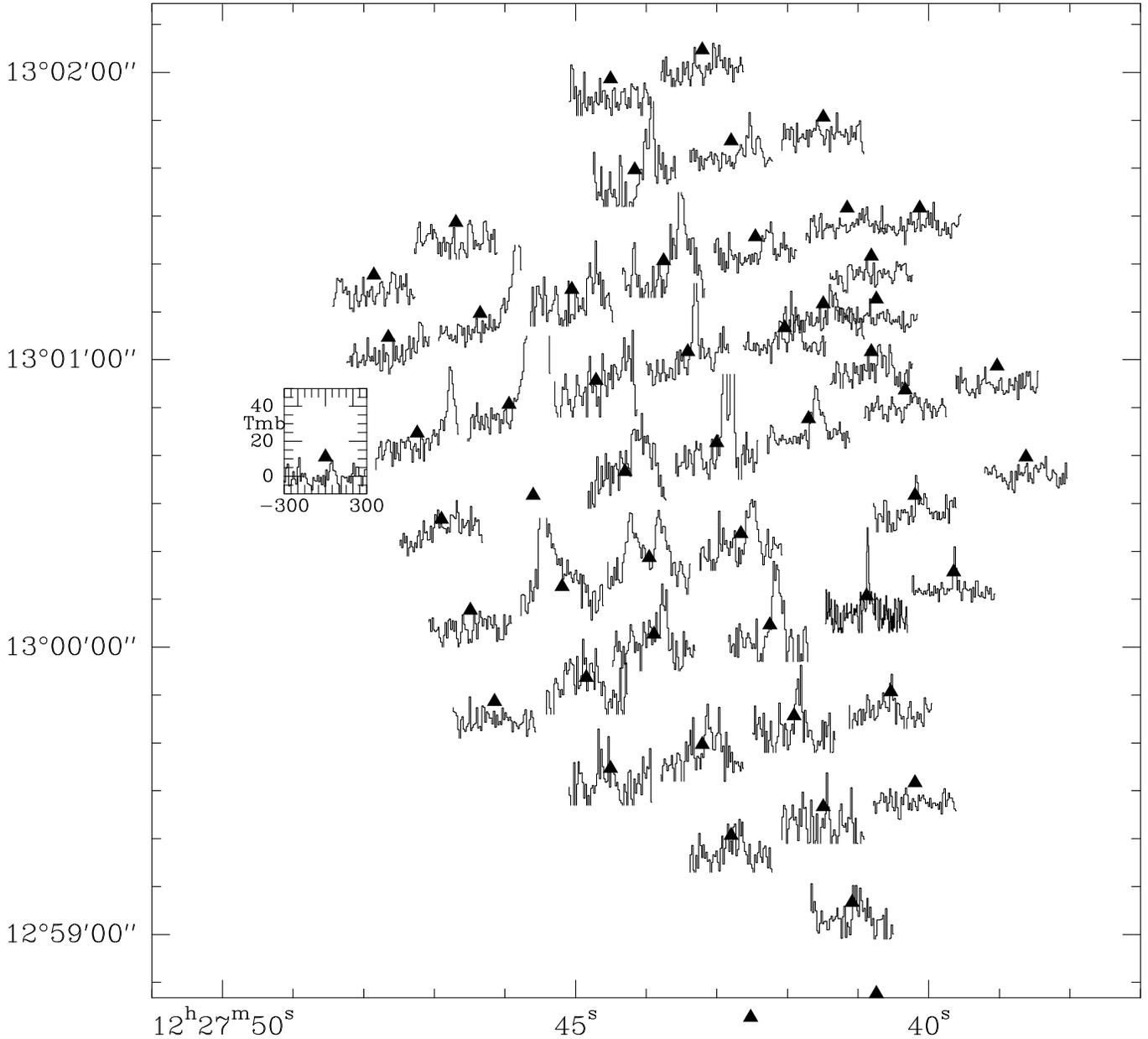,angle=-90,width=\hsize}
        \caption{Observed CO(1--0) spectra of the central part of NGC~4438
	on an optical B band image. The systemic velocity of NGC~4438 is
	70~km\,s$^{-1}$.
        } \label{fig:coobs}
\end{figure*}

The CO(1--0) spectra of NGC~4438 superimposed on an optical 
image are shown in Fig.~\ref{fig:coobs}.
We have omitted the central spectrum which we show separately in
Fig.~\ref{fig:windspec}. 

Using the same conversion factor as above, 
the total molecular gas mass (including the He associated
with the H$_{2}$) of NGC~4438 is $\sim 1.7 \times 10^{9}$~M$_{\odot}$.  
The conversion ratio is very probably overestimated in the central 
parts (as for spirals in general) and possibly underestimated in the outer
regions.  Our observations show weaker CO emission for the western part
than reported by Combes et al. (1988) and that it is closely related
(spatially) with the H{\sc i} and dust lanes.

The CO(1--0) emission extends over almost the entire observed
region. The lines are becoming weaker to the west. 
Whereas we detect only one line at
the eastern edge of the galactic disk north-east of the galaxy center, CO emission is
detected up to $\sim 80''$ to the west of the galaxy's major axis. Moreover, all
spectra south west of the galaxy center peak at velocities greater than zero, despite
the fact that the southern part represents the approaching side of the galaxy. 
This behaviour is also present in the H$\alpha$ spectra of Kenney et al. (1995).
We will show in Sect.~\ref{sec:rps} that these spectra are displaced due to the action
of ram pressure.  Most striking is the detection of strong, clearly separated double lines
west and south west of the galaxy center at the positions ($-20''$,$-20''$), ($-38''$, $10''$),
and ($-40''$, $-10''$). The spectrum south of the galaxy center ($-6''$, $-19''$) 
also shows a redshifted component which can be better seen in the H$\alpha$
Fabry-Perot data of Chemin et al. (2005). Such planar and extraplanar double lines are a unique feature, 
characteristic of the interaction NGC~4438 underwent and is still undergoing.

The extraplanar gas in NGC 4438, or gas at velocities other than rotation, 
is at least $4.7 \times 10^{8}$~M$_{\odot}$ when the He is included.  We took 
the positive velocity component of all spectra west of a right ascension offset of $-19''$.
The spectrum at an offset of $-19.5''$ was not included although its velocity has clearly been 
increased with respect to normal rotation.  Similarly, the spectra along the major axis 
south of the center have not been included although they too have velocities 
which are not those of normal rotation (or normal rotation and tides).

\subsection{The northern tidal tail \label{sec:ntt}}

\begin{figure}
        \resizebox{\hsize}{!}{\includegraphics{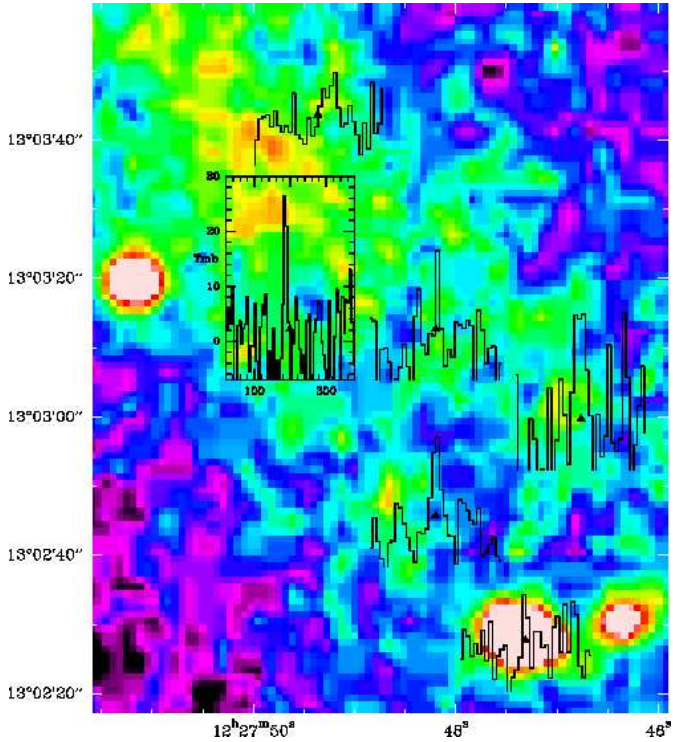}}
        \caption{Observed CO(1--0) spectra of the northern part of NGC~4438
	on an optical B band image. The systemic velocity of NGC~4438 is
	70~km\,s$^{-1}$.
        } \label{fig:north}
\end{figure}

As shown in Fig.~\ref{fig:optical} we also observed several positions on the northern
tidal tail (Fig.~\ref{fig:north}).
Unexpectedly, we detect CO(1--0) emission with narrow line profiles at several positions
($\Delta v \la 50$~km\,s$^{-1}$). Using the same conversion factor of 
$\ratio = 2 \times 10^{20}$~H$_2$~cm$^{-2}$ (K km/s)$^{-1}$, the molecular gas mass 
in this region is about $2.5 \times 10^{7}$~M$_{\odot}$.

\section{Simulations \label{sec:simulations}}

We have adopted a model where the ISM is simulated as a collisional component,
i.e. as discrete particles which have a mass and a radius and which
can have inelastic collisions (sticky particles).
In contrast to smoothed particle hydrodynamics (SPH), which is a 
quasi continuous approach and where the particles cannot penetrate each other, our approach 
allows a finite penetration length, which is given by the mass-radius relation of the particles.
Both methods have their advantages and their limits.
The advantage of our approach is that ram pressure can be included easily as an additional
acceleration on particles that are not protected by other particles (see Vollmer et al. 2001).
In this way we avoid the problem of treating the huge density contrast between the 
ICM ($n \sim 10^{-4}$~cm$^{-3}$) and the ISM ($n > 1$~cm$^{-3}$) of the galaxy.

Since the model is described in detail in Vollmer et al. (2001), we 
will summarize only its main features.
The N-body code consists of two components: a non-collisional component
that simulates the stellar bulge/disk and the dark halo, and a
collisional component that simulates the ISM.
The 25\,000 particles of the collisional component represent gas cloud complexes which are 
evolving in the gravitational potential of NGC~4438. For the ISM-ISM collision
models the ISM of NGC~4435 is simulated by 10\,000 particles.

The total assumed initial gas masses of NGC~4438 and NGC4435 are 
$M_{\rm gas}^{\rm tot}=4.9\,10^{9}$~M$_{\odot}$ and $1.3\,10^{9}$~M$_{\odot}$, respectively.
For NGC~4438 we begin with a NGC~4501 type galaxy, i.e. the initial H{\sc i}
diameter is close to the optical diameter. The recent GALEX observations of 
Boselli et al. (2005) justify this assumption (see also Sect.~\ref{sec:rps}).
A larger H{\sc i} diameter
does not change our results. The total gas mass of NGC~4438 calculated as 
the sum of the molecular gas mass derived from our observations and the initial H{\sc i}
mass of NGC~4501 gives $3.4 \times 10^{9}$~M$_{\odot}$.
The total gas mass for NGC~4435 represents the upper limit for
an S0 galaxy of its luminosity (Welch \& Sage 2003).
The ensemble of model clouds has a power law mass distribution as described in Vollmer et al. (2001).
A radius is attributed to each particle depending on its mass. 
During the disk evolution the particles can have inelastic collisions, 
the outcome of which (coalescence, mass exchange, or fragmentation) 
is simplified following Wiegel (1994). This results in an effective gas viscosity in the disk. 

As the galaxy moves through the ICM, its clouds are accelerated by
ram pressure. Within the galaxy's inertial system its clouds
are exposed to a wind coming from a direction opposite to that of the galaxy's 
motion through the ICM. 
The temporal ram pressure profile has the form of a Lorentzian,
which is realistic for galaxies on highly eccentric orbits within the
Virgo cluster (Vollmer et al. 2001).
The effect of ram pressure on the clouds is simulated by an additional
force on the clouds in the wind direction. Only clouds which
are not protected by other clouds against the wind are affected.

The non--collisional component of NGC~4438 consists of 65\,536 particles, which simulate
the galactic halo, bulge, and disk. NGC~4435 is modeled by 49\,152 particles.
The characteristics of the different galactic components are listed in
Table~\ref{tab:param}.
\begin{table}
      \caption{Total mass, number of particles $N$, particle mass $M$, and smoothing
        length $l$ for the different galactic components.}
         \label{tab:param}
      \[
         \begin{array}{lllll}
           \hline
           \noalign{\smallskip}
           {\rm component} & M_{\rm tot}\ ({\rm M}$$_{\odot}$$)& N & M\ ({\rm M}$$_{\odot}$$) & l\ ({\rm pc}) \\
           \hline
	   {\rm {\bf NGC~4438}} & & & & \\
           {\rm halo} & 1.4\,10$$^{11}$$ & 32768 & $$4.4\,10^{6}$$ & 1200 \\
           {\rm bulge} & 5.1\,10$$^{9}$$ & 16384 & $$3.1\,10^{5}$$ & 180 \\
           {\rm disk} & 2.5\,10$$^{10}$$ & 32768 & $$7.6\,10^{5}$$ & 240 \\
	   {\rm {\bf NGC~4435}} & & & & \\
	   {\rm halo} & 1.1\,10$$^{11}$$ & 16384 & $$6.8\,10^{6}$$ & 1200 \\
           {\rm bulge} & 1.7\,10$$^{10}$$ & 16384 & $$1.0\,10^{6}$$ & 180 \\
           {\rm disk} & 5.5\,10$$^{9}$$ & 16384 & $$3.4\,10^{5}$$ & 240 \\
           \noalign{\smallskip}
        \hline
        \end{array}
      \]
\end{table}
The total masses of NGC~4438 and NGC~4435 are $1.7\,10^{11}$~M$_{\odot}$ and $1.3\,10^{11}$~M$_{\odot}$,
respectively. 
We have chosen to model NGC~4435 as a bulge dominated system with a bulge to disk mass ratio of 3:1.
The bulge has an exponential surface brightness profile with a scale length of $2$~kpc,
which is the H band exponential scale length given by Gavazzi et al. (2003) ($R_{\rm e}=25''$).
The disk scale lengths are 3.5~kpc and 0.9~kpc for NGC~4438 and NGC~4435.
The assumed value for disk scale length of NGC~4438 is based on the optical diameter of another Virgo 
spiral of slightly higher luminosity, NGC~4501 ($D_{25}=6.9'=34$~kpc, and the
relation between the optical radius $R_{25}$ and the H band disk scale length 
$R_{\rm e}$ derived by Gavazzi et al. (2000): $R_{25} \sim 5 \times R_{\rm e}$.

The resulting flat rotation velocities of NGC~4438 and NGC~4435 are $\sim 160$~km\,s$^{-1}$ and
$\sim 250$~km\,s$^{-1}$. Our choice of the rotation velocities are close to those derived from
our observed CO line profiles (Sect.~\ref{sec:n4435} and \ref{sec:rps}).
The disk of NGC~4435 has an inclination angle of $\sim 70^{\circ}$.
NGC~4435 has a higher rotation velocity
despite its smaller mass, because its disk scale length is smaller than that of NGC~4438.
For comparison, Combes et al. (1988) used a mass ratio between NGC~4438 and NGC~4435 of 2:1
and a ratio between the disk scale lengths of 4:1. 
The particle trajectories are integrated using an adaptive timestep for
each particle. This method is described in Springel et al. (2001).
The following criterion for an individual timestep is applied:
\begin{equation}
\Delta t_{\rm i} = \frac{20~{\rm km\,s}^{-1}}{a_{\rm i}}\ ,
\end{equation}
where $a_{i}$ is the acceleration of the particle i.
The minimum value of $t_{\rm i}$ defines the global timestep used 
for the Burlisch--Stoer integrator that integrates the collisional
component.

We present 4 different simulations:
\begin{enumerate}
\item
tidal interaction with NGC~4435 only,
\item
tidal interaction and ISM-ISM collision between NGC~4438 and NGC~4435,
\item
tidal interaction and ram pressure stripping,
\item
tidal interaction, ISM-ISM collision and ram pressure stripping.
\end{enumerate}
For the simulations without an ISM-ISM collision, NGC~4435 is modeled only
by a non-collisional component (halo, bulge and disk).

In a different set of simulation we assumed NGC~4435 to be a disk dominated system before
the interaction. These simulations show that (i) NGC~4435 cannot be transformed into
a bulge dominated system by the tidal interaction and (ii) the tidal effects on
NGC~4438 are qualitatively and quantitatively the same as for the simulations
with an initially bulge dominated NGC~4435.

\subsection{ISM-ISM collision \label{sec:ismism}}

In a widely accepted picture (see, e.g. Kulkarni \& Heiles 1988,
Spitzer 1990, McKee 1995), the ISM of the Galaxy consists of 5 different phases:
the molecular, cold neutral, warm neutral, warm ionized, and hot ionized gas phases.
In our case, the molecular and neutral gas phases are of interest.

We have assumed that in the inner 10~kpc of a spiral galaxy,
half of the neutral gas mass is molecular whereas the other half is in atomic form.
Moreover, deep H{\sc i} observations of local spiral galaxies (Braun 1997) showed that
60-90\% of the H{\i} emission is in form of cold atomic hydrogen.
In the calculation of the mass fractions we use a cold to total H{\sc i} fraction of 70\%.
More than 80\% of the total gas mass within 10~kpc is neutral and more than 70\% has temperatures 
well below 1000~K.

Two gas phases are important for an ISM-ISM collision: the warm ($T \sim 10^{4}$~K), diffuse 
(ionized or not) and the cold ($T \la 100$~K), dense phases, because they represent more than 90\,\%
of the total gas mass. The diffuse and dense phases have volume filling factors of 
$\Phi_{\rm V} \sim$0.3-0.5 and 0.02, respectively. 
The area filling factor for the diffuse phase is between 0.5 and 1 and that of the dense phase is
about 0.1 (Braun 1997). However, only 20\% of the total mass is warm but 75\% is cold.
Our simulations do not distinguish between the two phases.
Therefore, we use average volume and area filling factors of
$\Phi_{\rm V}=\big( \Sigma_{i} \frac{4 \pi}{3} r_{{\rm cl}, i}^{3} \big)/V_{\rm gal} \sim 0.05$ 
and $\Phi_{\rm A}= \big( \Sigma_{i} \pi r_{{\rm cl}, i}^{2} \big)/A_{\rm gal} \sim 0.2$ in our ISM-ISM collision 
simulations, which takes the different volume filling factors and mass fractions into account,
where $r_{\rm cl}$ is the cloud radius and $V_{\rm gal}$/$A_{\rm gal}$ are the volume/area occupied
by the gas cloud complexes. The collision rate can be controled by changing the
mass-radius relation to obtain clouds of larger radii, i.e. larger cross section.
The most massive, i.e. the largest clouds determine the volume filling factor.
The chosen volume filling factor insures that each cloud of
NGC~4435 undergoes at least one collision during the ISM-ISM collision
between NGC~4438 and NGC~4435. 
We have run an additional set of simulations where we increased the cloud-cloud collision
rate by a factor of 4. The higher collision rate does not change the distribution and
dynamics of the high density gas (CO) significantly. However, the dynamics of the low
density gas located in the north-western low surface brightness stellar tidal tail are different.
Since we do not have observations in this area, it is not possible to discriminate between the
simulations with different collision rates.

In general, SPH simulations which include heating and cooling (Barnes \& Hernquist 1996, Struck 1997,
Tsuchiya et al. 1998) show that cooling is very important (mainly because it is proportional to
the square of the gas density) and that cooling times are so small that in general
the gas can be regarded approximately as isothermal. This is what we assume in our model.
In Sect.~\ref{sec:tidalism} we show that our results are similar to those of SPH ISM-ISM
collision simulations of Struck (1997) and Tsuchiya (1998), 
which treat the gas as a continuous phase ($\Phi_{\rm V} \sim 1$).

We now estimate the mass of the warm phase involved in the ISM-ISM collision between
NGC~4438 and NGC~4435: since the fraction of the NGC~4438 disk, which is hit by NGC~4435,
is about 10\% of the total disk area, the mass of the warm ISM of NGC~4438 involved in the interaction
is $0.1 \times 0.2 \times$M$_{\rm gas}^{\rm tot}$. Moreover, the total warm gas mass
of NGC~4435 is involved in the interaction: $0.2 \times $M$_{\rm gas}^{\rm tot}$.
Thus even asuming that all clouds in NGC~4435
experience a collision with clouds in NGC~4438, only $3.4\times 10^{8}$~M$_{/odot}$ due to 
the collision of the two warm phases is affected, an unknown fraction of which
could be found between the two galaxies. If, in addition, there is an efficient interaction
between the warm and the cool phases, this value can increase up to $\sim 1.5 \times 10^{9}$~M$_{\odot}$.
This interaction between a warm, diffuse phase of one galaxy
with the cold, dense phase of the other galaxy is not taken into account in our model.
We discuss the implications of this estimate in Sect.~\ref{sec:tidalism}.

\subsection{Ram pressure \label{sec:rpps}}

For the simulations including ram pressure stripping we assume that the 
galaxy is on an eccentric orbit within the cluster. The temporal
ram pressure profile can be described by:
\begin{equation}
p_{\rm ram}=p_{\rm max} \frac{t_{\rm HW}^{2}}{t^{2}+t_{\rm HW}^{2}}\ ,
\end{equation}
where $t_{\rm HW}$ is the width of the profile (Vollmer et al. 2001). 
We can estimate the maximum ram pressure using the formula of
Gunn \& Gott (1972):
\begin{equation}
2 \pi G \Sigma_{*} \Sigma \simeq v_{\rm rot}^{2} R^{-1} \Sigma \simeq p_{\rm ram}\ ,
\end{equation}
where $G$ is the gravitational constant, $\Sigma_{*}$ the stellar surface density.
The truncation radius of the gas disk of NGC~4438 of $R$=2.5~kpc together with a
rotation velocity of $v_{\rm rot}=160$~km\,s$^{-1}$ and a gas surface density
of $\Sigma=1.5 \times 10^{21}$~cm$^{-2}$ implies a maximum ram pressure of 
$p_{\rm max}$=5000~cm$^{-3}$(km/s)$^{2}$. A realistic orbit in the Virgo cluster
then implies $t_{\rm HW}$=50~Myr. In order to produce extraplanar, high gas 
column density gas, the inclination angle between the disk and the orbital plane must
be higher than $\sim 30^{\circ}$. On the other hand, 
the efficiency of ram pressure depends on the inclination angle $i$
between the galactic disk and the orbital plane (Vollmer et al. 2001).
Too low inclination angles
(more face-on stripping) result in a very high column density gas in the south of
the galaxy center. The azimuthal angle of the galaxy's motion within the
ICM with respect to the orbit of NGC~4435 is chosen so as to reproduce the asymmetric 
CO and H$\alpha$ emission distributions.
In the end, we set $p_{\rm max}$=5000~cm$^{-3}$(km/s)$^{2}$, $t_{\rm HW}$=50~Myr,
and an inclination angle of $i$=68$^{\rm o}$. This resulted in the best model fit
to observations. For ram pressure stripping the model clouds have a column density of
$\Sigma=1 \times 10^{21}$~cm$^{-2}$.

\subsection{Relative importance of the interactions \label{sec:relimp}}

In order to investigate the importance of the different interactions, 
we estimate the accelerations on the ISM of NGC~4438 due to the galactic
potential ($a_{\rm gal}$), the direct gravitational pull of NGC~4435 
which is close to the tidal acceleration ($a_{\rm tid}$), 
and ram pressure ($a_{\rm ram}$):
\begin{equation}
a_{\rm gal} \sim \frac{M_{\rm N4438} G}{R^{2}}\ ,
\end{equation}
\begin{equation}
a_{\rm tid} \sim \frac{M_{\rm N4435} G}{(r-R)^{2}}\ ,
\end{equation}
\begin{equation}
a_{\rm ram} \sim \frac{p_{\rm ram}}{\Sigma_{\rm ISM}}\ ,
\end{equation}
where $G$ is the gravitational constant, $v_{\rm rot,N4438}$, $v_{\rm rot,N4435}$ are the
rotation velocities of NGC~4438 and NGC~4435, respectively, $R$ is the radius of tidal influence,
$r$ is the distance between the two galaxies, $p_{\rm ram}$ is the ram pressure,
and $\Sigma_{\rm ISM}$ is the gas surface density of NGC~4438 at the radius $R$. 
For the tidal interaction to be important the following condition must be fulfilled:
\begin{equation}
\frac{a_{\rm tid}}{a_{\rm gal}} = \frac{M_{\rm N4435}}{M_{\rm N4438}} \frac{R^{2}}{(r-R)^{2}} \sim 1\ .
\end{equation}
The mass fraction between the two galaxies is about 0.75.
With an impact parameter of $r=8.5$~kpc the parts of NGC~4438 at radii $R > 4$~kpc are affected.
For ram pressure to be important we obtain the following condition:
\begin{equation}
\frac{a_{\rm ram}}{a_{\rm gal}} = \frac{p_{\rm ram} R \cos i}{\Sigma_{\rm ISM} v_{\rm rot, N4438}^{2}} \sim 1\ ,
\end{equation}
where $i$ is the angle between the two vectors $\vec{a_{\rm ram}}$ and $\vec{a_{\rm gal}}$.
A peak ram pressure of $p_{\rm ram}=5000$~cm$^{-3}$(km\,s$^{-1}$)$^{2}$, a rotation velocity
of $v_{\rm rot, N4438}=160$~km\,s$^{-1}$, $i=0^{\circ}$, and a radius
of $R=4$~kpc results in a critical gas surface density of $\Sigma_{\rm ISM, crit} \sim 2
\times 10^{21}$~cm$^{-2}$.
Thus, ISM with surface densities smaller than $\Sigma_{\rm ISM, crit}$ is stripped
by ram pressure at radii greater than 4~kpc. This is close to the observed 
stripping radius of $\sim 50''$ (Sect.~\ref{sec:neutral}). On the other hand, giant 
molecular clouds have column densities much higher than $10^{21}$~cm$^{-2}$
(several $10^{22}$~cm$^{-2}$) and thus should not be stripped. 

A single simulation requires about 100 hours on a 1.2~GHz and 1024~MB RAM memory PC.
In the following we describe the temporal evolution of the 4 different simulations.
The models presented here are those with the best results (i.e. best set of collision
parameters, see Sect.~\ref{sec:rpps} and \ref{sec:ismism}) for each of the four scenarios.

\subsection{Tidal interaction only \label{sec:tidalo}}

For the tidal interaction we follow Combes et al. (1988) and chose a close, rapid, retrograde
encounter between NGC~4438 and NGC~4435. The closest approach occurs at $t=165$~Myr.
The impact parameter is 9~kpc and the velocity difference is $\Delta v=840$~km\,s$^{-1}$.
We verified their best fit model in varying the impact parameters
from 5 to 10~kpc and the inclination angle between the orbital plane of NGC~4435 and the
disk plane of NGC~4438 between $20^{\circ}$ and $40^{\circ}$.

\begin{figure*}
   \resizebox{\hsize}{!}{\includegraphics{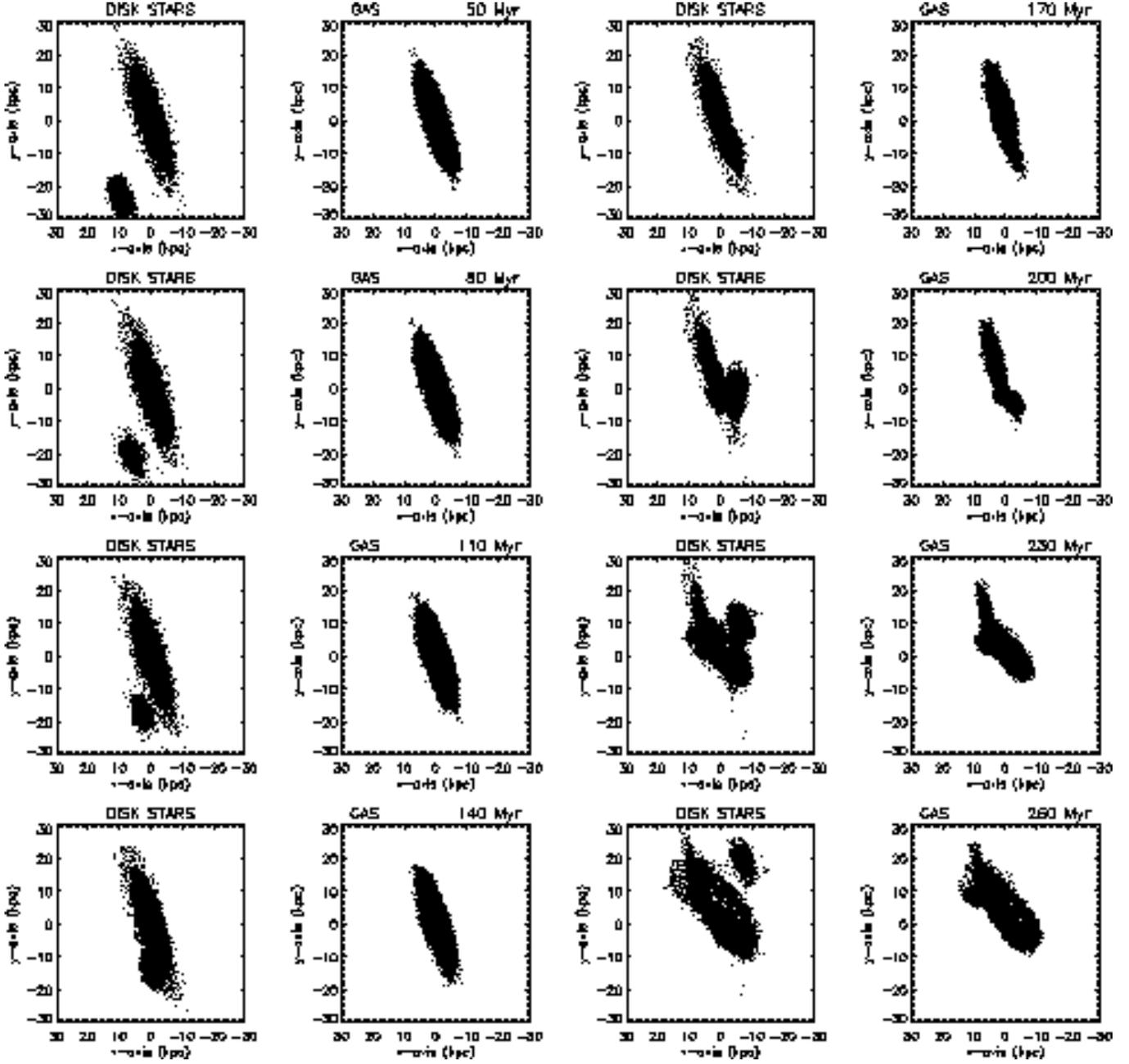}}
        \caption{Tidal interaction only: evolution of the model stellar (1st 
	and 3rd columns) and gaseous disks (2nd and 4th columns) of NGC~4438. 
	The major axis position angle and inclination of NGC~4438 
	are $PA=20^{\rm o}$ and $i=80^{\rm o}$, respectively.
	For clarity, we only show the stellar disk of NGC~4435.
	The elapsed time from the beginning of the simulations
	is shown above the panels showing the ISM.
	NGC~4435 passes through the disk of NGC~4438 at $t=165$~Myr.
	A movie of this simulation can be found here.
        } \label{fig:tidalo}
\end{figure*}

There is no significant change in the distribution of stars and gas of NGC~4438
before the passage of NGC~4435 at $t=165$~Myr. 
Being more extended, NGC~4438 gets most damaged during the interaction. A prominent tidal tail is formed
north of the galaxy center. In addition, a second tidal tail in the south forms
a loop whose northern/western edge moves towards the north/west while its southern/eastern
edge does not change position. There is no stellar or gas bridge between NGC~4438 and
NGC~4435. At the end of the simulation ($t=260$~Myr) the distribution of gas and stars
of NGC~4438 is highly asymmetric with a sharply truncated southern and eastern edge, a prominent
tidal arm towards the north, and an extended component to the west (Fig.~\ref{fig:tidalo}).

NGC~4435 is located close to its observed location. The difference in radial velocities
of NGC~4438 and NGC~4435 is $\Delta v_{\rm r}=670$~km\,s$^{-1}$, compatible with the
observed value of $\sim 730$~km\,s$^{-1}$.
The final stellar distribution at the timestep which we compare to our CO observations
is discussed in Sect.~\ref{sec:final}.

\subsection{Tidal interaction and ISM-ISM collision \label{sec:tidalism}}

\begin{figure*}
  \resizebox{\hsize}{!}{\includegraphics{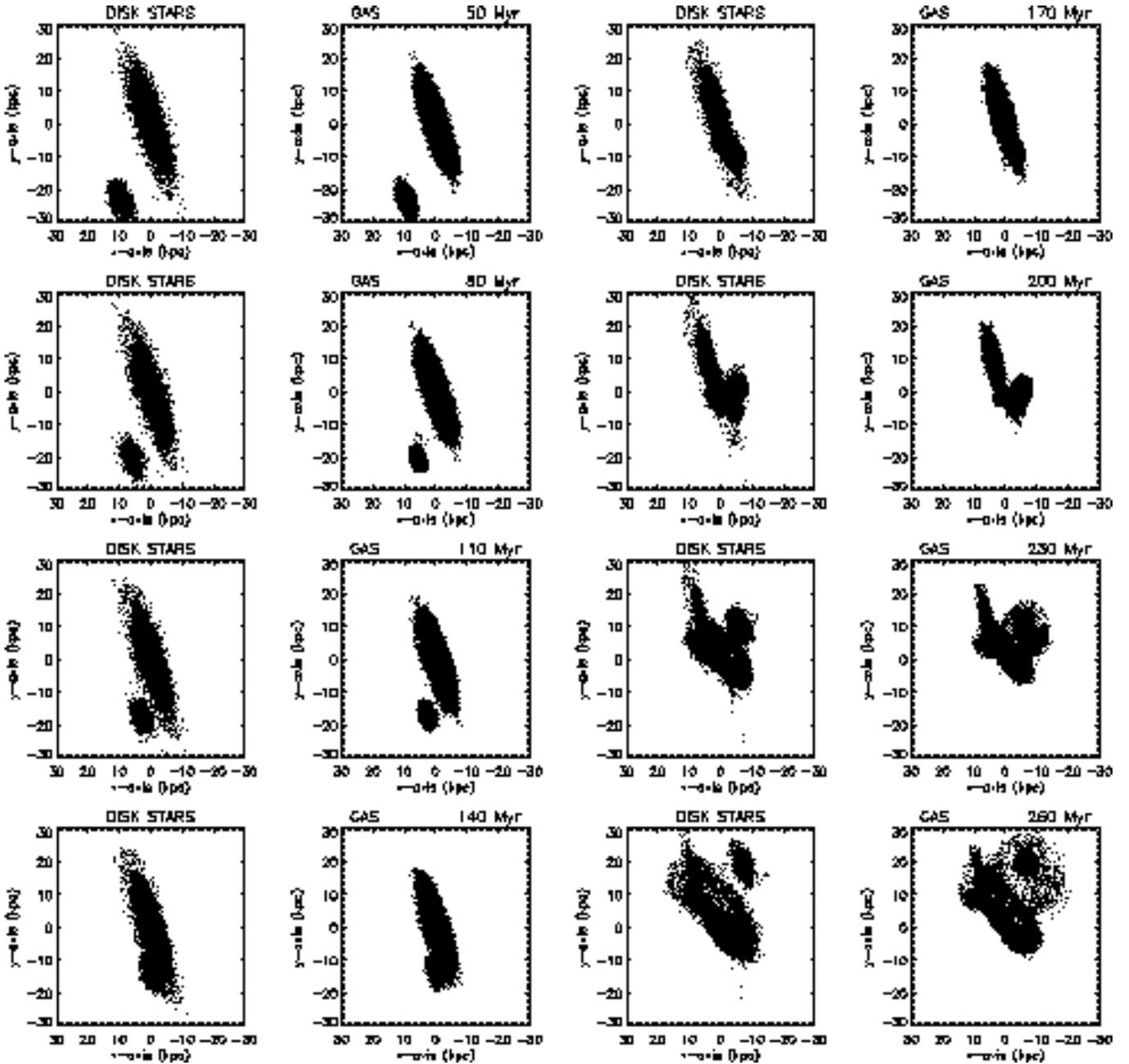}}
        \caption{Tidal interaction and ISM-ISM collision: 
	evolution of the model stellar (1st and 3rd columns)
	and gaseous disks (2nd and 4th columns). 
	The major axis position angle and inclination of NGC~4438 
	are $PA=20^{\rm o}$ and $i=80^{\rm o}$, respectively.
	For clarity, we only show the stellar disk of NGC~4435.
	The elapsed time from the beginning of the simulations
	is shown above the panels showing the ISM.
	NGC~4435 passes through the disk of NGC~4438 at $t=165$~Myr.
	A movie of this simulation can be found here.
        } \label{fig:tidalism}
\end{figure*}

Since the collision between the ISM of NGC~4438 and NGC~4435 only concerns the gaseous
component whose mass is small compared to that of the non-collisional component,
the evolution of the stellar disk is the same as for the simulations of the tidal interaction alone
(Sect.~\ref{sec:tidalo}). After the ISM-ISM collision at $t=165$~Myr, two gaseous bridges are
formed between NGC~4438 and NGC~4435 (Fig.~\ref{fig:tidalism}). 
The southern arm is more prominent and longer lasting than the northern component.
The total gas mass in the gas bridge is about $3.4 \times 10^{8}$~M$_{\odot}$ which
represents $\sim$25\% of the gas mass of NGC~4435. 
The surface density within the whole bridge is high (5-10~M$_{\odot}$pc$^{-2}$) until 45~Myr after the
collision ($t=165$~Myr). Still $\sim 20$~Myr later, the surface density in the middle of the bridge drops
to $\sim 3$~M$_{\odot}$pc$^{-2}$ and higher surface density gas (5~M$_{\odot}$pc$^{-2}$)
is only found along the minor axis at distances smaller than 4~kpc to the west of the disk of NGC~4438.
At $t=260$~Myr (95~Myr after the collision) this close, western extraplanar gas has only a mean surface 
density of about 2~M$_{\odot}$pc$^{-2}$.
The rest of the gas in the bridge has a lower surface density.
At the end of the simulation ($t=260$~Myr) the gas distribution
of NGC~4438 is very similar to that of the simulations of the tidal interaction alone, i.e.
the ISM-ISM collision is neither able to remove the gas from the northern tidal tail, nor
can it produce extraplanar, high surface density gas $\sim 100$~Myr after the collision.

Our simulations can be compared to the SPH ISM-ISM collision simulations of Struck (1997)
and Tsuchiya et al. (1998). Since SPH particles cannot interpenetrate by definition, the intruder
galaxy makes a hole in the primary galaxy, which is rapidly filled due to differential
rotation (Struck 1997). Whereas Struck (1997) explains the effects in more detail, 
Tsuchiya et al. (1998) give more quantitative information. In particular, Tsuchiya et al. (1998)
show the evolution of a head-on collision of a primary galaxy with a massive companion in their Fig.~6.
The relative velocity is $300$~km\,s$^{-1}$, thus a factor 3 smaller than in our case and their
companion mass is 25\% of the primary galaxy. They find 25\% of the mass of the companion's
gas mass in the bridge. In addition, the surface density of the 
gas in the bridge drops drastically within 40~Myr (from $T=15=195$~Myr to $T=18=234$~Myr) 
and $\sim 70$~Myr after the collision the gas surface density in the bridge is small. 
The column density evolution of the gas in the bridge is qualitative and, as far as we can see, 
in quantitative agreement with
our findings. Interestingly, position-velocity plots of their simulations at
$T=15$ or $T=18$ (their Fig.~9) show that
the gas in the bridge near the primary galaxy has negative velocities with respect to the
primary's systemic velocity, because the bridge gas is falling back onto the primary galaxy.

If an ISM-ISM collision were responsible for the observed extraplanar CO, more than $10^{8}$~M$_{\odot}$
of cold and dense gas must have been involved in the collision, since the involved diffuse 
ISM is not massive enough to account for the observed extraplanar, molecular gas mass 
(see Sect.~\ref{sec:ismism}).
It is not likely that the extraplanar gas still has a high surface density
($\geq 10$~M$_{\odot}$pc$^{-2}$) as it is observed $\sim 100$~Myr after the collision.
Furthermore, $\sim 100$~Myr after the collision, the gas which was involved in the collision
should be falling back onto NGC~4438 producing lines at negative velocities in the south-west
of the galaxy center, which is 
not observed. We conclude that an ISM-ISM collision is not responsible for the
observed molecular gas distribution and kinematics.

\subsection{Tidal interaction and ram pressure stripping}

\begin{figure*}
  \resizebox{\hsize}{!}{\includegraphics{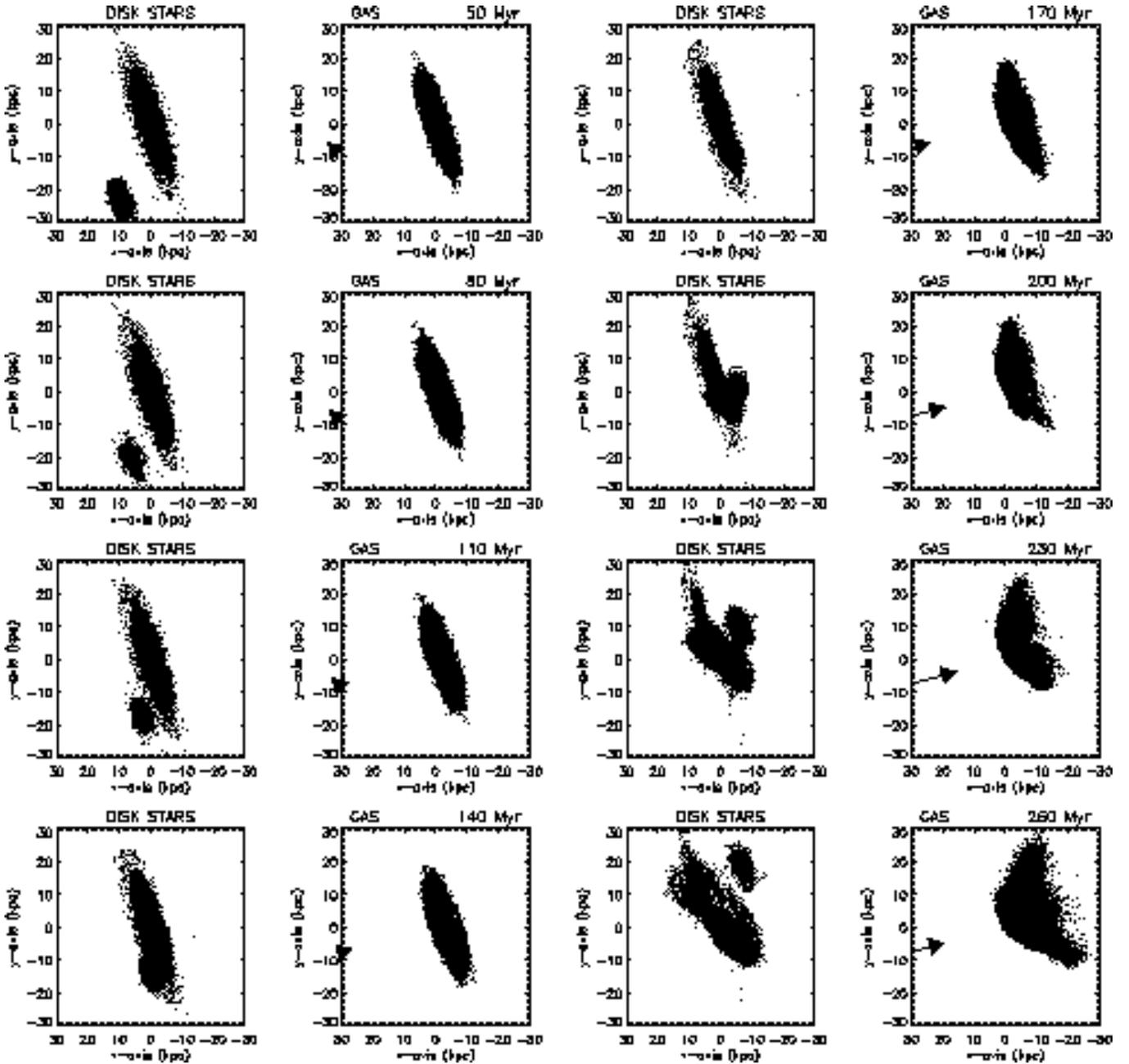}}
        \caption{Tidal interaction and ram pressure stripping:
	evolution of the model stellar (1st and 3rd columns)
	and gaseous disks (2nd and 4th columns). 
	The major axis position angle and inclination of NGC~4438 
	are $PA=20^{\rm o}$ and $i=80^{\rm o}$, respectively.
	The elapsed time from the beginning of the simulations
	is shown above the panels showing the ISM.
	NGC~4435 passes through the disk of NGC~4438 at $t=165$~Myr.
	The arrow indicates the direction of ram pressure, i.e. it is opposite to
	the galaxy's velocity vector, and its size is proportional
	to $\rho v_{\rm gal}^{2}$. The galaxy passes the cluster core at
	250~Myr (maximum ram pressure).
	For clarity, we only show the stellar disk of NGC~4435.
	A movie of this simulation can be found here.
	A movie of the simulation including a tidal interaction,
	an ISM-ISM collision, and ram pressure can be found here.
        } \label{fig:tidalrps}
\end{figure*}

In this simulation we did not assign gas to NGC~4435 but include ICM ram pressure.
The choice of the free parameters of the ISM-ICM interaction are given in Sect.~\ref{sec:rpps}.
Since ram pressure acts only on the ISM of NGC~4438, the evolution of the stellar disk is the same as 
for the other simulations (Sect.~\ref{sec:tidalo} and \ref{sec:tidalism}).
For the gas, the situation changes completely. Soon after the passage of NGC~4435, the
gas distribution is distorted by the action of ram pressure, which pushes the ISM of
NGC~4438 to the west (Fig.~\ref{fig:tidalrps}).
The probability for a close galaxy encounter is highest in the cluster core
due to its strongly peaked density of elliptical and S0 galaxies (Schindler et al. 1999).
In addition, in the cluster core the galaxy velocity and the ICM density, i.e. ram pressure, 
are highest. Thus the small time difference between the galaxy collision and maximum 
ram pressure of $\Delta t=85$~Myr is a natural consequence of the maximized probability for
galaxy collisions and maximized ram pressure in the cluster core. The direction of
ram pressure is natural for a highly eccentric orbit of NGC~4438 in the Virgo cluster
(Vollmer et al. 2001). The trajectory of NGC~4435 can also be naturally explained by
a less eccentric orbit.

\subsection{Tidal interaction, ISM-ISM collision and ram pressure stripping \label{sec:tidalismrps}}

This simulation contains all three interactions: (i) the tidal interaction, (ii) the ISM-ISM
collision and (iii) ram pressure stripping.
The evolution of the gaseous component is very similar to that of the previous simulation
(tidal interaction and ram pressure stripping), because ram pressure stripping is the 
most energetic and thus most important effect.
Due to the plot mode (each particle is represented by a black dot), the evolutionary plot of
this simulation is indistinguishable from that of Fig.\ref{fig:tidalrps}.
However, the differences can be seen in the animation or in the final gas distribution
(Sect.~\ref{sec:final}).

\subsection{The final stellar and gas distributions \label{sec:final}}

\begin{figure*}
  \resizebox{\hsize}{!}{\includegraphics{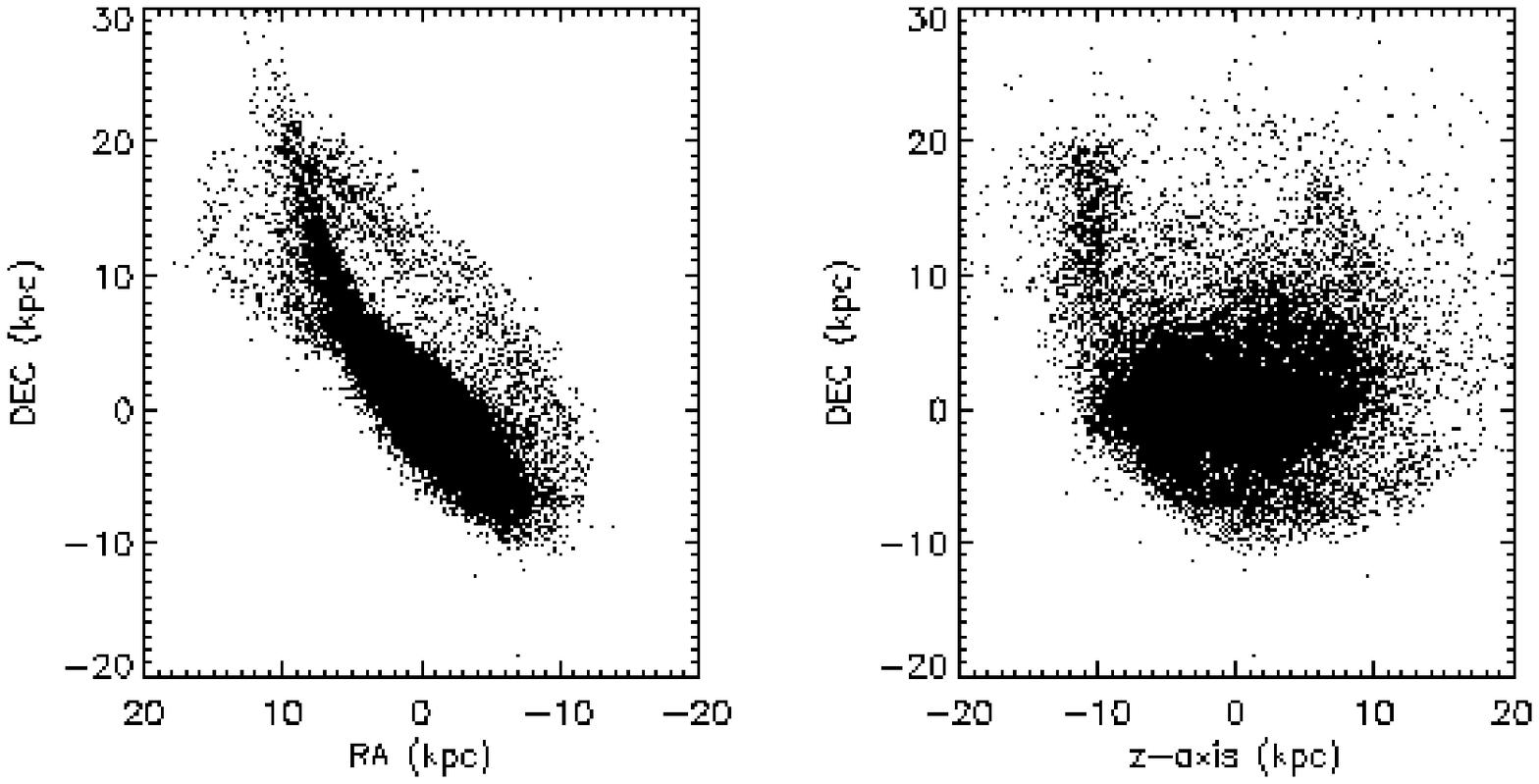}}
  \caption{Final snapshot of the stellar content at $t=260$~Myr. Left panel: 
      projected with the position and inclination angle of NGC~4438. Right panel:
      Perpendicular view.
        } \label{fig:stars}
\end{figure*}

The final distribution of the stellar content of NGC~4438 and NGC~4435 is shown in Fig.~\ref{fig:stars}.
The overall distribution of NGC~4438 is close to the observed one (Fig.~\ref{fig:optical}), i.e. the major 
characteristics are reproduced: (i) the stellar disk is truncated to the south,
(ii) there is a prominent northern tidal arm and (iii) displaced stellar debris to the west
of the galaxy's main disk.  In addition, the location and radial
velocity of NGC~4435 is close to observations (Sect.~\ref{sec:tidalo}).
The reproduction of the stellar 
component is imperfect in that the model arm to the North is offset.
Whereas the observed arm is located on the
galaxy's major axis, the model arm is parallel to the major axis, but $\sim 3$~kpc offset to the east. 
The aim here is not to present a perfect model of the tidal interaction but to 
study the influence of the different kinds of interactions, for which the model stellar
distribution is amply sufficient.

We produced a model cube of the gas within the inner $20 \times 20$~kpc of our final snapshot.
From this model cube we extracted column density maps and model spectra at different resolutions.
Since our model gas distribution of NGC~4438 does not include the high density core that is observed,
we added this component ad hoc. If we had included this component into the numerical simulations,
the computational costs would have been too high to carry out our systematic study.
The a posteriori addition of this core, does not change our conclusions, because the core
does not extend further out than the model gas distribution without the core at $t=260$~Myr. 
The core mainly enhances
the surface brightness of the central disk. In addition, the initial model rotation curve
of NGC~4438 is not as steep as it is observed because of numerical resolution. The additional
core component takes this into account.

\begin{figure}
        \resizebox{\hsize}{!}{\includegraphics{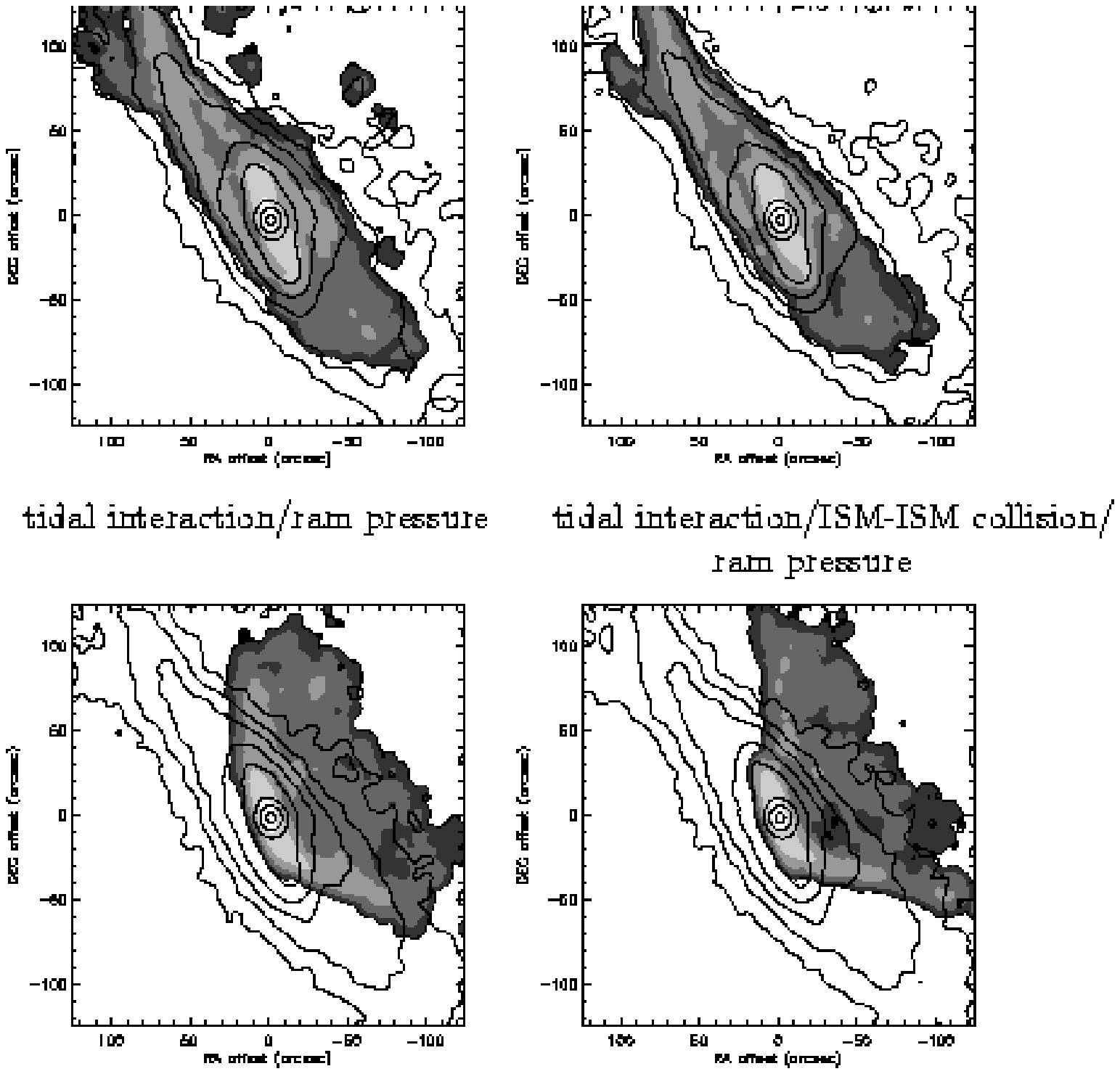}}
        \caption{Simulations of NGC~4438. Contour: stellar distribution.
	Greyscale: gas distribution at a resolution of $7''$. 
	The dark regions correspond to low surface density gas.
        } \label{fig:plotall}
\end{figure}

An overlay of the stellar component with the gas distribution convolved to a 
resolution of $7''=580$~pc for all four simulations is shown in Fig.~\ref{fig:plotall}.
The lower panels include ram pressure while the upper panels do not.
The final gas distributions of the simulations including ram pressure are very different
from those without ram pressure, whereas the effect of the ISM-ISM collision on the
gas distribution is minor. This shows that ram pressure is much more important
for the overall gas dynamics than the ISM-ISM collision.
The gas distributions of the simulations of the tidal interaction with and without an
ISM-ISM collision are similar. Both distributions show two tidal arms in the north.
Moreover, high column density gas is found to the west
of the galactic disk, which belongs to the second spiral arm extending to the south-west.
Combes et al. (1988) already observed this effect.

The two simulations including ram pressure stripping also lead to similar 
distributions. The ISM-ISM collision enhances the velocity dispersion and
decreases the volume density of NGC~4438's gas which is involved in the collision.
Due to the decrease of the particle density the ICM penetration length into the ISM of 
NGC~4438 increases and thus the ram pressure efficiency increases. The net effect is that
more gas is pushed to the west by ram pressure when an additional ISM-ISM collision occurs.
The gaseous disk of NGC~4438 is truncated at a radius of $\sim 40''$ and the
gas of the outer disk is pushed to the west. The displaced gas of highest surface
density is found in the south-west and north-west close to the truncated disk (see also
Sect.~\ref{sec:neutral}).

\subsection{Distribution of the neutral ISM \label{sec:neutral}}

\begin{figure}
        \resizebox{\hsize}{!}{\includegraphics{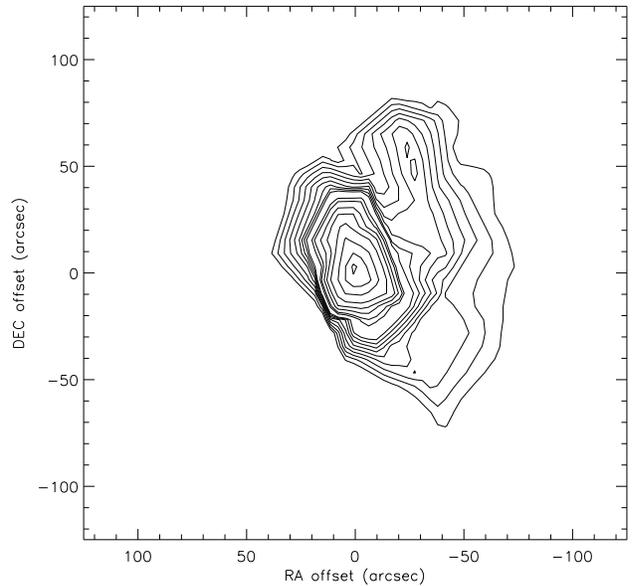}}
        \caption{Column density map of the CO(1--0) observations.
        } \label{fig:mom0obs}
\end{figure}

Since the pointings of our observations form an irregular grid, we used a Delaunay triangulation
within IDL (TRIANGULATE and TRIGRID functions, see the IDL reference guide) 
to interpolate the data and to derive a column density map (Fig.~\ref{fig:mom0obs}).
Within the galactic disk we observe an asymmetry along the major axis, i.e. there
is more flux in the north than in the south. The western, extended emission shows the
same asymmetry, i.e. there is a local maximum in the north. This emission closely traces
the dust absorption features observed towards the west of the galactic disk
of NGC~4438 (see e.g. Fig.~1 of Kenney et al. 1995). 

\begin{figure}
        \resizebox{\hsize}{!}{\includegraphics{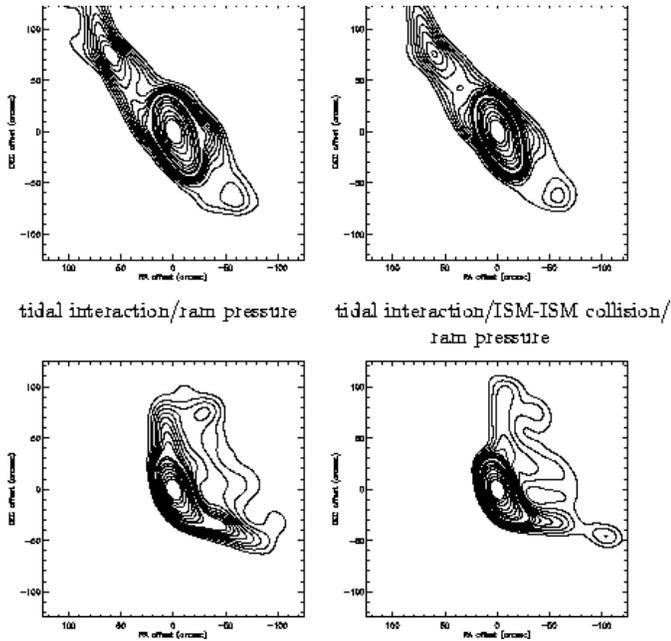}}
        \caption{Column density maps of the simulations. The contour levels are the same
	as for Fig.~\ref{fig:mom0obs}.
        } \label{fig:mom0all}
\end{figure}

The column density maps of our 4 simulations are shown in Fig.~\ref{fig:mom0all}.
Their spatial sampling is complete. For the comparison with the observed column density map
one has to keep in mind that these are irregularly gridded and do not cover the
entire region.
Only the simulations that include ram pressure stripping reproduce the observed CO
gas distribution. In particular, the simulations without ram-pressure
cannot account for the observed strong CO(1--0) lines west of the
galaxy center. The simulation including all interactions (tidal, ISM-ISM collision and
ram pressure stripping) is also consistent with our observations. The main difference
with the tidal/ram pressure simulation is that the column density of the western 
extraplanar gas is lower if an ISM-ISM collision is included, because the gas clouds 
extracted and ``heated'' from NGC~4438 during this ISM-ISM collision are efficiently stripped by 
ram pressure.

The main difference between our CO(1--0) observations and the simulations including
ram pressure stripping is that the model gas surface density is greater than observed
in the southern, extraplanar gas. Since this is not observed in CO, we suspect that
part of this gas is heated and ionized with its molecular gas dissociated.
A part of this ``missing'' gas might be visible in X-rays or in the radio continuum.

\section{The need for ram pressure \label{sec:rps}}

We showed in Sect.~\ref{sec:neutral} that only 
simulations including ram pressure can account for the fact that there is
almost no neutral gas detected in the northern tidal tail. Boselli et al. (2005)
observed NGC~4438 with the GALEX satellite. They found UV emission in the northern 
and southern tidal tail where no gas is detected. By fitting a spectral energy distribution
locally using NIR, optical and UV data they reconstructed the star formation history
of different regions. For the northern and southern tidal tail the data is consistent
with a star burst $\geq 100$~Myr ago. This is entirely consistent with our scenario
for the tidal interaction. Since there is UV radiation
detected in the tidal tails, gas was associated with these regions before the interaction,
which justifies our initial gas distribution of NGC~4438.

In the following we show that the observed CO(1--0) spatial and velocity asymmetry and extraplanar double 
lines (Fig.~\ref{fig:coobs}) can be reproduced by a model that includes ram pressure stripping
(Fig.~\ref{fig:data221}).
We recall that, in order to reproduce the observed stellar distribution, 
we adopted the orbital parameters
of Combes et al. (1988). We verified their best fit model in the way described in Sect.~\ref{sec:tidalo}.
With these parameters, the parameters for the tidal interaction and the ISM-ISM collision are fixed.
Concerning ram pressure, the open parameters are the maximum ram pressure,
the inclination angle between the disk and the orbital plane and the azimuthal projection angle 
(Vollmer et al. 2001; see also Sect.~\ref{sec:rpps}).
The maximum ram pressure is chosen such that there is extraplanar high column density gas as observed.
The inclination angle is chosen such that there is a a minimum of
gas to the south-west. When we decreased the inclination angle, there was far too much
extraplanar gas in the SW. The azimuthal projection angle is chosen such
that the position and radial velocity of NGC~4435 are close to observations. We call this model our best fitting model.

\subsection{The best fitting model: tidal interaction and ram pressure stripping \label{sec:doubles}}

\begin{figure}
	\psfig{file=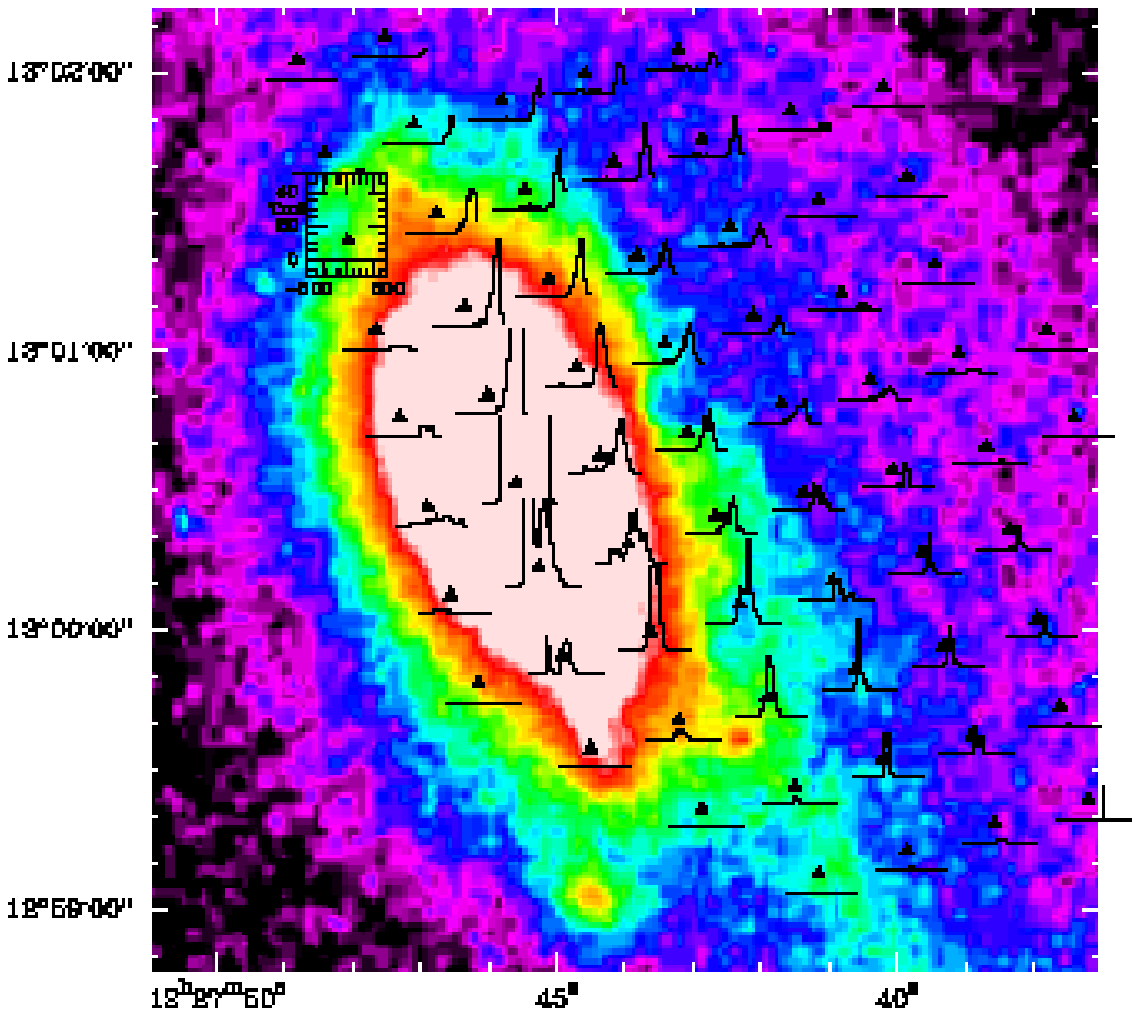,angle=-90,width=\hsize}
	\psfig{file=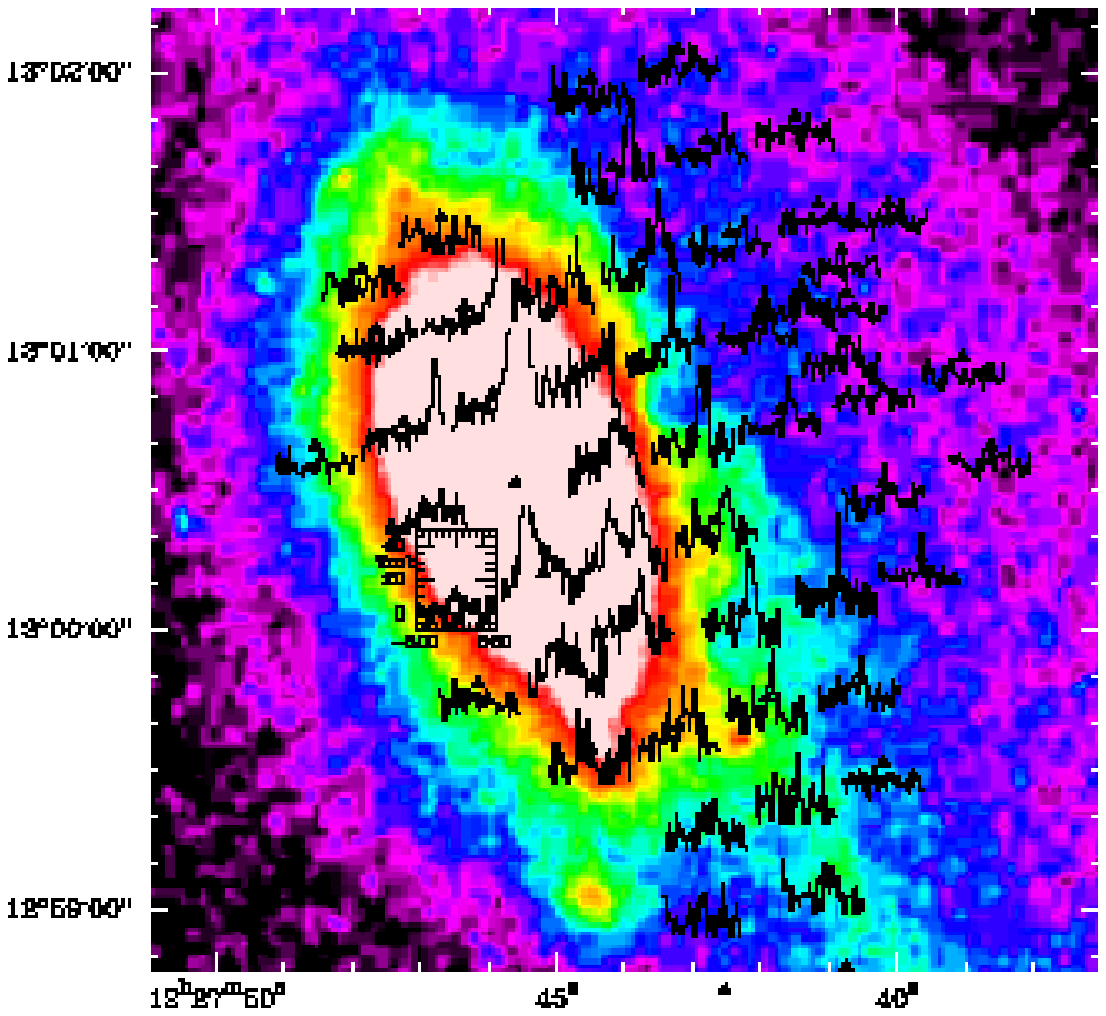,angle=-90,width=\hsize}
        \caption{Top panel: tidal interaction and ram pressure stripping.
	Simulated CO(1--0) spectra of the central part of NGC~4438
	on an optical B band image. The systemic velocity of NGC~4438 is
	70~km\,s$^{-1}$.
	Bottom panel: Observed $^{12}$CO(1-0) spectra (Fig.~\ref{fig:coobs}). 
        } \label{fig:data221}
\end{figure}

We took the last snapshot of Fig.~\ref{fig:tidalrps} and produced a model cube out of it
(see also Sect.~\ref{sec:final}). CO model spectra with a resolution of $21''$ were then
extracted on a equidistant grid with a cell size of $20''$. We only show model spectra with a 
non zero flux (Fig.~\ref{fig:data221}).  
The model spectra show the same overall east-west asymmetry as is observed
(see also Sect.~\ref{sec:neutral}).
At the eastern edge of the galactic disk the model shows weak lines that
are consistent with our CO(1--0) observations. In particular, the observed spectrum 
about $30''$ north-east of the galaxy center is well reproduced, even though the maximum of the 
model spectrum is weaker than the observed one. We find strong model lines to the
north-west of the galaxy center as observed. The model spectra in the south-west of the galaxy peak
at velocities greater than zero as observed. In addition, the model also reproduces the observed 
double line profile of the south-western spectrum. Whereas the two observed peaks have the same fluxes, 
the model peak at high velocities is much stronger than the peak at low velocities.
The double line profile is still visible in the model spectra further to the west.
The two lines trace two spatially distinct regions in the line of sight. The
line at negative velocities traces gas which rotates in the galactic disk, whereas the line
at positive velocities traces gas which is accelerated by ram pressure.

\begin{figure}
        \resizebox{\hsize}{!}{\includegraphics{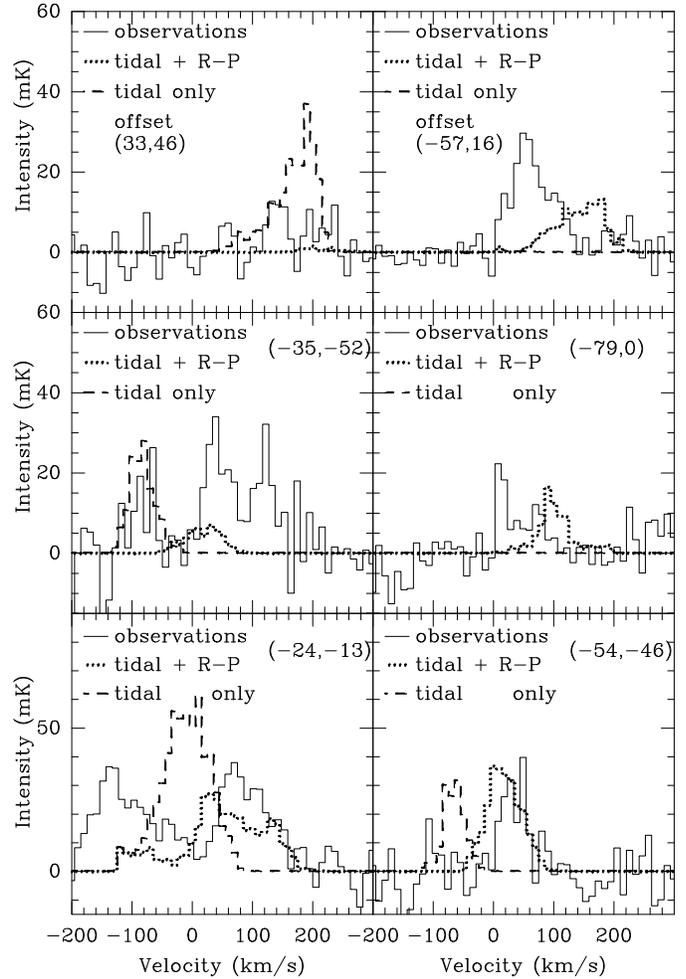}}
        \caption{Representative spectra to the NE, SW, and West in NGC~4438
	  shown to illustrate generic properties of the difference between 
	  collisions with and without ram pressure stripping (R-P).  
	  Offset in arcseconds with respect to the center of NGC~4438 are 
	  indicated in each panel. Solid lines: observations. Dotted lines:
	  tidal interaction and ram pressure stripping. Dashed lines:
	  tidal interaction only.
	} \label{fig:newfig}
\end{figure}

For more detailed comparison we show selected observed and model spectra in 
Fig.~\ref{fig:newfig}.
The observed spectra (solid line) follow the tidal and ram pressure simulations 
(dotted line) more closely than the tidal interaction alone (dashed line).
Adding the ISM-ISM collision has little effect.
On both sides of the galaxy the gas velocities are {\it not} well 
represented  by the tidal only simulations whereas they are 
reasonably reproduced when ram pressure is included.  Molecular gas is present 
in many positions to the West where the tidal forces do not bring gas
but is not present on or near the northern or eastern part of the
main body of NGC~4438, where strong lines are expected if ram pressure is
not acting.  NGC~4435 penetrates NGC~4438 in a region which has now 
rotated to the northern side.  A "smooth" or clumpy ISM-ISM collision 
$\sim 100$~Myr ago could not change the rotation velocities in the southern part on
or near the major axis ({\it i.e.} pos. $-35''$,$-52''$) but the observations 
show that to the South the CO velocities are much higher than they 
would be in the absence of ram pressure.
Most of the molecular gas has been efficiently stripped from what is now the 
northern part of NGC~4438.  Only the simulations with ram pressure show this 
(see also Fig.~\ref{fig:mom0all}) and a collision with the ISM (dense or diffuse)
of NGC~4435 could not empty a large region of its dense gas.  
While the intensity ratio is not reproduced, it is very interesting that at
the ($-24''$, $-13''$) position the ram pressure simulation reproduces the velocities
whereas almost no gas is at the velocity of the big (single) peak 
predicted by the tidal interaction only.  The ($-43''$, $-8''$) position is similar.
Here as well, the ISM-ISM collision has little effect.

The three spectra to the right of Fig.~\ref{fig:newfig} are all considerably off the plane of 
NGC~4438 to the west. The simulations without ram pressure show no flux here or,
for the ($-54''$, $-46''$) position, at a significantly lower velocity as in the 
spectra to the right.

\begin{figure}
        \resizebox{\hsize}{!}{\includegraphics{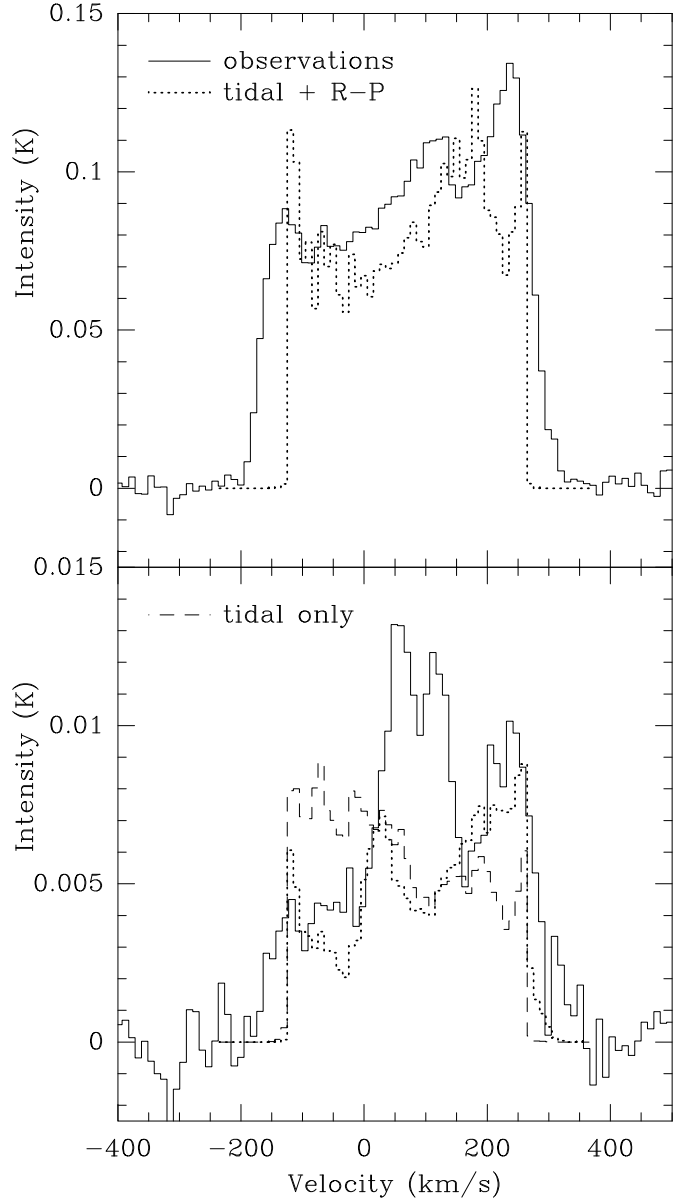}}
        \caption{Upper panel: central spectrum. Lower panel: integrated spectrum
	of the entire galaxy.
	Solid line: observed spectrum. Dotted line: model spectrum of the simulation
	including a tidal interaction and ram pressure stripping. Dashed line:
	model spectrum of the simulation including a tidal interaction alone.
	The systemic velocity of NGC~4438 is 70~km\,s$^{-1}$.
        } \label{fig:windspec}
\end{figure}

The observed central and integrated spectra together with the model spectra are shown
in Fig.~\ref{fig:windspec}.
In the model spectrum the double horn component at high absolute velocities in the 
central spectrum is mainly due to the fast rotating core of NGC~4438
(see Sect.~\ref{sec:final}). On the other hand, the bump at $v \sim 140$~km\,s$^{-1}$ is
due to gas that has been accelerated by ram pressure to positive velocity. 
Thus, we conclude that the observed bump at 
$v \sim 100$~km\,s$^{-1}$ is due to ram pressure stripping. The simulated
bump is shifted by $\sim 40$~km\,s$^{-1}$ with respect to the observed bump.

The integrated spectrum shows two main characteristics: (i) a major bump
around the galaxy's systemic velocity $v=70$~km\,s$^{-1}$ and (ii) an 
asymmetry with less gas on the low velocity (approaching) side.
The simulations including only a tidal interaction (dashed line) show neither a central bump
nor the observed asymmetry in velocity. On the contrary, the model integrated spectrum
without ram-pressure shows a reversed asymmetry with more gas at low velocities, 
because most of the spectra summed are on the approaching side.
On the other hand, the integrated spectrum of the simulations including a tidal
interaction and ram pressure stripping do reproduce the observed asymmetry and
and the central bump. However, the model bump has the same peak value as the
high velocity peak and is shifted to smaller velocities.

We conclude that the model including a tidal interaction and ram pressure stripping
reproduces the major characteristics of the observed spectra:
\begin{itemize}
\item
the east west asymmetry,
\item
the displacement of the peaks of the spectra to higher velocities in the south-west of the galaxy,
\item
the double line profiles located to the west and south-west of the galaxy center.
\end{itemize}
Since the detection of extraplanar 
CO shows that molecular gas is displaced, this implies that either the bulk of the molecular gas is taken
along with the atomic gas via ram pressure
due to the diffuse ISM or magnetic field coupling, or the molecular clouds are left behind,
form stars rapidly and then are destroyed by the energy input due to star formation. This means that the
Gunn \& Gott criterion might not be applicable to single clouds, but only to the ISM within the 
galactic disk as a whole.

\subsection{Alternative scenarios \label{sec:alternative}}

We have chosen the model including a tidal interaction and ram pressure stripping
as the best fitting model, although including an ISM-ISM collision is about as good. 
In this section we present the alternative scenarios
(tidal interaction alone; tidal interaction and ISM-ISM collision; tidal interaction,
ISM-ISM collision and ram pressure stripping).

\begin{figure}
	\psfig{file=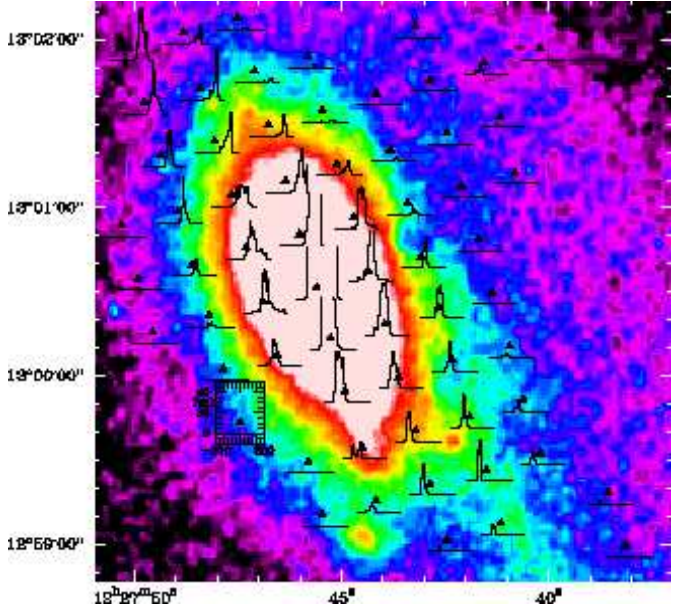,angle=-90,width=\hsize}
        \caption{Tidal interaction only. Simulated CO(1--0) spectra of the central part of NGC~4438
	on an optical B band image. The systemic velocity of NGC~4438 is
	70~km\,s$^{-1}$.
        } \label{fig:data217}
\end{figure}

The spectra from the simulation including the tidal interaction alone are shown in
Fig.~\ref{fig:data217}.
High column density gas is dragged to the west by the tidal interaction. However, 
the spectra with non zero flux do not extend further to the west than $60''$ (see also Fig.~\ref{fig:mom0all}).
Contrary to observations, there are strong lines at the eastern edge of the galactic disk, but not to the
north-west of the galaxy center. All spectra located
south of the galaxy center peak at velocities smaller than zero in contrast to the observed 
spectra in this region. At the north-eastern edge of the optical disk multiple
line profiles can be seen. These are due to the superposition of the disk gas and gas located in the
north-western tidal arm (see Fig.~\ref{fig:mom0all}). There are no double line profiles in
the south-west of the galaxy.

\begin{figure}
	\psfig{file=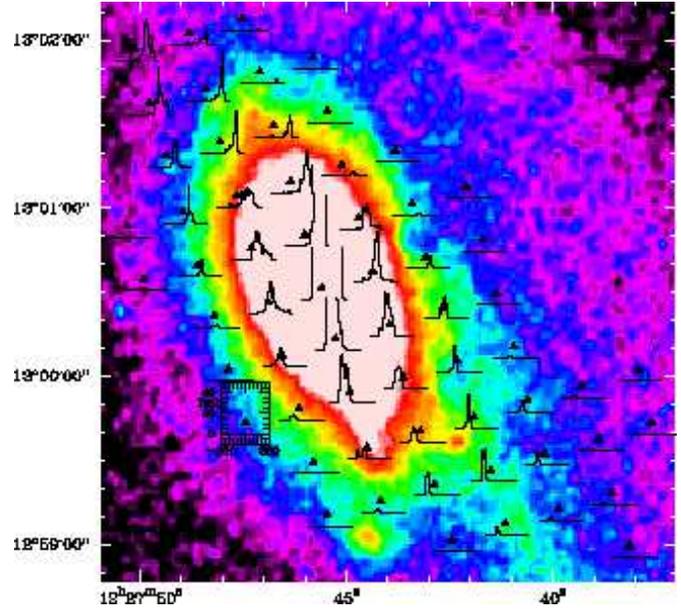,angle=-90,width=\hsize}
        \caption{Tidal interaction and ISM-ISM collision. 
	Simulated CO(1--0) spectra of the central part of NGC~4438
	on an optical B band image. The systemic velocity of NGC~4438 is
	70~km\,s$^{-1}$.
        } \label{fig:data223}
\end{figure}

The spectra from the simulation adding ISM-ISM collision are shown in
Fig.~\ref{fig:data223}.
The spectra within and east of the optical disk are very similar to those of the
simulation with a tidal interaction alone (Fig.~\ref{fig:data217}).
The spectral profiles at and beyond the western edge of the optical disk occupy the
same velocity range as those of Fig.~\ref{fig:data217}, but their peak fluxes are
lower. In contrast to the observed spectra, the model spectra in the south-west 
of the galaxy center show peaks at negative velocities.

\begin{figure}
	\psfig{file=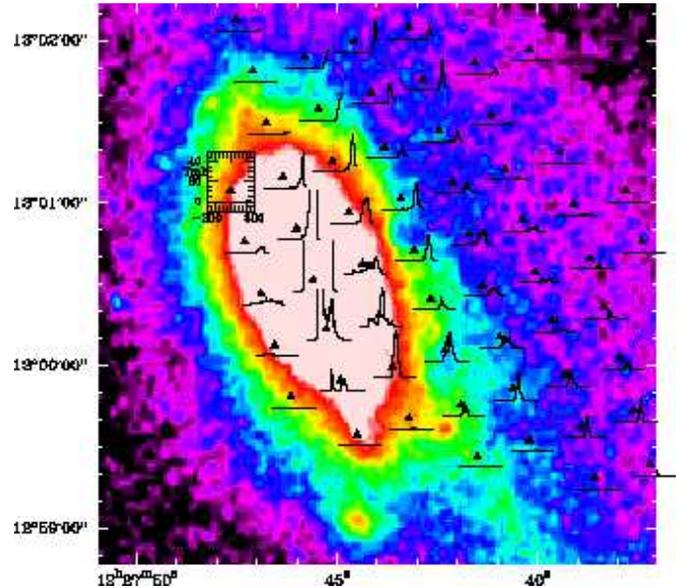,angle=-90,width=\hsize}
        \caption{Tidal interaction, ISM-ISM collision and ram pressure stripping. 
	Simulated CO(1--0) spectra of the central part of NGC~4438
	on an optical B band image. The systemic velocity of NGC~4438 is
	70~km\,s$^{-1}$.
        } \label{fig:data224}
\end{figure}

The spectra from the simulation including the tidal interaction, an ISM-ISM collision and
ram pressure stripping is shown in Fig.~\ref{fig:data224}.
The spectra are very similar to those of the simulation including a tidal interaction and
ram pressure stripping. Again, the effect of the ISM-ISM collision is minor compared to that of the
tidal interaction.

We conclude that the simulations without ram pressure do not reproduce the major 
characteristics of the CO(1--0) observations: (i) the spectra at the eastern edge of the optical disk, 
(ii) the strong lines to the north-west of the galaxy center, 
(iii) the double line profile in the south-west of the galaxy center, and (iv) the shift of the
line profile in the south-west to positive velocities. Only the simulations with ram
pressure stripping are able to reproduce these features.

\section{The northern tidal tail \label{sec:north}}

We have detected $2 \times 10^{7}$~M$_{\odot}$ of molecular gas in the northern tidal tail
and the mass limit for atomic gas in this region is a few $10^{7}$~M$_{\odot}$ (Hibbard et al. 2001).
The simulations including ram pressure stripping do not show any gas in or near this region
(see Fig.~\ref{fig:tidalrps}) nor is any H{\sc i} present. 
The gas mass in the northern tidal arm region of the simulations without ram-pressure
(Fig.\ref{fig:data217}) is $\sim 7\,10^{8}$~M$_{\odot}$.
Thus, the observed molecular gas mass given the absence of H{\sc i} in this region represents 
a few percent of the gas mass that would be there without ram pressure stripping.
This is consistent with a picture where the gas, which was in the form of molecular clouds
during the phase of maximum ram pressure ($t > 150$~Myr), was not affected by ram pressure
due to its high gas surface density. This implies that the observed molecular clouds
in the northern tidal arm region are rather long-lived (several 10~Myr).

\begin{figure}
	\psfig{file=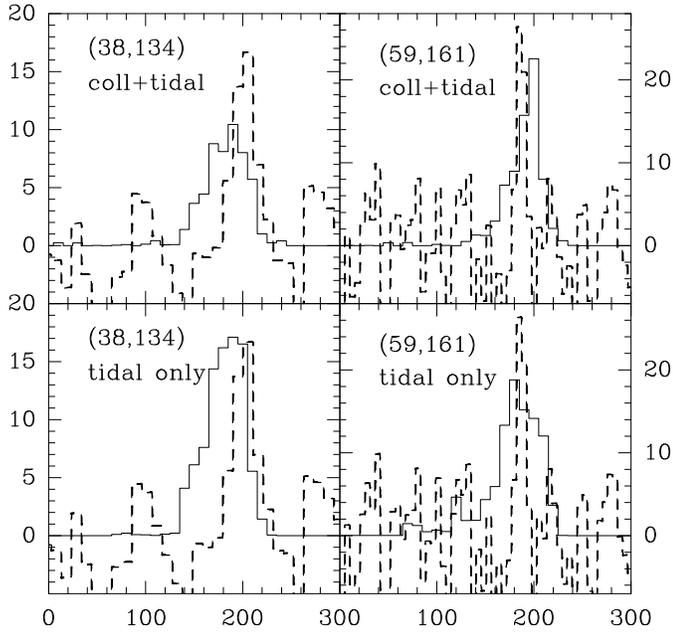,angle=-90,width=\hsize}
        \caption{Observed (dashed line) CO(1--0) and simulated (solid line) spectra of 
	the northern part of NGC~4438. Upper left panel: tidal interaction and
	ISM-ISM collision. Lower panels: tidal interaction only.
	The position offsets in arcsec with respect to the galaxy center
	are shown in the upper left corner of each panel.
	The x-axis represent the radial velocity in km\,s$^{-1}$, the
	y-axis the main beam antenna temperature in mK.
	The systemic velocity of NGC~4438 is 70~km\,s$^{-1}$.
        } \label{fig:northmodel}
\end{figure}

Fig.~\ref{fig:northmodel} shows model and observed
spectra of two positions within the northern tidal arm region. 
The model spectra are rescaled to the show a maximum comparable to that
of the observed spectra. Note, however, that the mass in the model
spectrum is more than 30 times higher than the mass derived from the observed spectra.
Since the model northern tidal
arm is $\sim 3$~kpc offset from the galaxy's major axis to the east, the positions of the model
spectra are also shifted by 3~kpc to the east.   
Indeed, the model spectra show the same peak velocity and the same line width. 
The simulation including an ISM-ISM collision fits the observations at ($38''$, $134''$)
slightly better, because the secondary peak seen in the simulations of the tidal interaction alone
is less present.

Thus some giant molecular clouds (a few percent of the total gas mass) seem to have decoupled
from the ram pressure wind. They were left behind in the otherwise gas free disk and
survived for several 10~Myr. We cannot say if this is a phenomenon exclusively related to the tidal
tail or if giant molecular clouds in other parts of the disk also decoupled from the wind. 
Our observations indicate that the bulk of the molecular gas is stripped by ram pressure.
Since only a few percent of the total gas mass decouple, decoupled gas would only represent
$\sim 7 \times 10^{7}$~M$_{\odot}$.

\section{Conclusions: the history of NGC~4438 \label{sec:history}}

New $^{12}$CO(1--0) observations of the NGC~4438/4435 system are presented.
For the first time CO is detected in NGC~4435. As already shown by Combes et al. (1988),
the distribution of molecular gas is highly truncated within the disk of
NGC~4438 and we find an extraplanar component up to $\sim 1.5'$ to the west
of the galaxy center. Within this extraplanar molecular gas we find double line profiles 
at distances up to $40''$ to the west and south-west of the galaxy center.
In addition  the lines in the south of NGC~4438 are all redshifted with respect to
galactic rotation. We argue that asymmetry of the molecular gas distribution, 
the double line profiles and the redshifted lines are characteristic for ram pressure stripping
of NGC~4438.

The combination of our new CO(1--0) observations with detailed numerical simulations
leads to the following interaction scenario for the NGC~4438/NGC~4435 system:
NGC~4435 passed through the disk of NGC~4438  $\sim 100$~Myr ago at a radial distance of
$\sim 5-10$~kpc. The encounter was rapid ($\Delta v \sim 830$~km\,s$^{-1}$) and retrograde
(see also Combes at al. 1988). With an impact parameter $< 10$~kpc an ISM-ISM
collision is unavoidable. Its importance depends on the initial gas distributions
in NGC~4435 and NGC~4438. The estimated extent of NGC~4435's observed gas disk is $\la 1$~kpc.
In our simulations the gas disk of NGC~4438 had an initial extent of $\sim 10$~kpc. Even 
with this initial extent the influence of an ISM-ISM collision on the final
gas distribution and velocities is small compared to that of ram pressure stripping. 
NGC~4438 evolves on an eccentric orbit within the Virgo cluster. We observe the galaxy
$\sim 10$~Myr after its closest passage to the cluster center (M87).
Our model infers a total velocity of $v_{\rm tot} \sim 2000$ of NGC~4438 with respect
to the cluster mean. The galaxy is located at a total distance of $\sim 350$~kpc from
the cluster center. With the assumed maximum ram pressure the ICM density
at the position of NGC~4438 is $n_{\rm ICM} \sim 10^{-3}$~cm$^{-3}$ which is high
but compatible with the ICM density at the distance of NGC~4438 derived from X-ray 
observations (Schindler et al. 1999).
Ram pressure plays a key role for the evolution of the gaseous component of NGC~4438 together
with the tidal interaction. The displacement of the line profiles to higher velocities
in the south-western region of the galaxy, the lack of CO emission in the eastern
optical disk, and the presence of double line profiles in the south-west of the galaxy center
are clear signs of ram pressure stripping.
There is evidence that in the northern tidal arm region a few percent of the ISM survived
ram-pressure stripping in the form of dense molecular clouds (Sect.~\ref{sec:north}). 
These clouds must be stable for several 10~Myr.
Vollmer et al. (2001) claimed that gas clouds located within the ICM
can be stable and are almost entirely molecular, i.e. they are selfgravitating, heated by the X-ray background 
and cooled by C{\sc ii} and O{\sc i} line emission. With a typical column density of 
about $10^{22}$~cm$^{-2}$ these clouds resist ram pressure. We suggest that the observed molecular clouds in the
northern tidal arm regions are of this kind.

NGC~4438 has been greatly affected by both the tidal interaction with NGC~4435 and 
ram pressure stripping due to the rapid motion of NGC~4438 through the intracluster medium.
It is difficult, however, to be more precise than we have been
about the relative roles of tides and ram pressure given their
partial degeneracy and the uncertainties in the simulations and observations.

\begin{acknowledgements}
Based on IRAM observations. IRAM is supported by INSU/CNRS (France), MPG (Germany), and IGN (Spain). 
We made use of a DSS image. The Digitized Sky Survey was produced at the Space Telescope Science Institute 
under U.S. Government grant NAG W-2166. The images of these surveys are based on photographic data obtained 
using the Oschin Schmidt Telescope on Palomar Mountain and the UK Schmidt Telescope. The plates were
 processed into the present compressed digital form with the permission of these institutions.
This research has made use of the GOLD Mine Database (Gavazzi et al. 2003). 
\end{acknowledgements}

\end{document}